\documentclass[]{aa}

\usepackage{graphicx}
\usepackage{txfonts}
\usepackage{subfigure}
\usepackage{natbib}
\usepackage{bm}
\usepackage{hyperref}

\usepackage{amstext}

\begin{document}

\newcommand{\hoA}{o-H$_2^{18}$O 1$_{10}$--1$_{01}$}
\newcommand{\hoB}{o-H$_2^{17}$O 1$_{10}$--1$_{01}$}
\newcommand{\hoC}{o-H$_2$O 1$_{10}$--1$_{01}$}
\newcommand{\hoD}{p-H$_2$O 2$_{11}$--2$_{02}$}
\newcommand{\hoE}{p-H$_2$O 2$_{02}$--1$_{11}$}
\newcommand{\hoF}{p-H$_2^{18}$O 2$_{02}$--1$_{11}$}
\newcommand{\hoG}{o-H$_2^{18}$O 3$_{12}$--3$_{03}$}
\newcommand{\hoH}{o-H$_2$O 3$_{12}$--3$_{03}$}
\newcommand{\hoI}{p-H$_2^{18}$O 1$_{11}$--0$_{00}$}
\newcommand{\hoJ}{p-H$_2^{17}$O 1$_{11}$--0$_{00}$}
\newcommand{\hoK}{p-H$_2$O 1$_{11}$--0$_{00}$}
\newcommand{\hoL}{o-H$_2$O 2$_{21}$--2$_{12}$}
\newcommand{\hoM}{o-H$_2^{17}$O 2$_{12}$--1$_{01}$}
\newcommand{\hoN}{o-H$_2$O 2$_{12}$--1$_{01}$}
\newcommand{\hoO}{p-H$_2^{18}$O 3$_{13}$--3$_{20}$}
\newcommand{\hoP}{p-H$_2$O 5$_{24}$--4$_{31}$}
\newcommand{\kms}{km~s$^{-1}$}
\newcommand{\ms}{m~s$^{-1}$}
\newcommand{\cms}{cm~s$^{-1}$}
\newcommand{\halp}{H$\alpha$}
\newcommand{\msun}{$\rm M_{\odot}$}
\newcommand{\etal}{et~al.~}
\newcommand{\vsini}{$v~sin~i$}
\newcommand{\ctemp}{$^{\circ}$C}
\newcommand{\ktemp}{$^{\circ}$K}
\newcommand{\be}{\begin{equation}}
\newcommand{\ee}{\end{equation}}
\newcommand{\bd}{\begin{displaymath}}
\newcommand{\ed}{\end{displaymath}}
\newcommand{\bi}{\begin{itemize}}
\newcommand{\ei}{\end{itemize}} 
\newcommand{\bfig}{\begin{figure}}
\newcommand{\efig}{\end{figure}}
\newcommand{\bc}{\begin{center}}
\newcommand{\ec}{\end{center}}
\newcommand{\hii}{{H\scriptsize{II}}}
\newcommand{\vlsr}{V$_{\mathrm{LSR}}$}
\newcommand{\vtur}{V$_{\textrm{\tiny{tur}}}$}
\newcommand{\vexp}{V$_{\textrm{\tiny{exp}}}$}
\newcommand{\vinfall}{V$_{\textrm{\tiny{infall}}}$}
\newcommand{\coa}{$^{12}\mathrm{CO}$}
\newcommand{\cob}{$^{13}\mathrm{CO}$}
\newcommand{\coc}{$\mathrm{C}^{18}\mathrm{O}$}
\newcommand{\lsun}{L$_{\odot}$~}
\newcommand{\lfir}{L$_{\textrm{\tiny{FIR}}}$}
\newcommand{\agua}{$X_{\textrm{\tiny{H$_2$O}}}$}
\newcommand{\ratioop}{o$/$p}
\newcommand{\ratiosept}{$X_{\textrm{\tiny{$^{18}$O$/$$^{17}$O}}}$}
\newcommand{\ratiohuit}{$X_{\textrm{\tiny{$^{16}$O$/$$^{18}$O}}}$}
\newcommand{\watersept}{the H$_2^{17}$O}
\newcommand{\waterhuit}{H$_2^{18}$O}
\newcommand{\water}{H$_2^{16}$O}
\newcommand{\lsol}{L$_\odot$\,}
\newcommand{\Msol}{M$_\odot$\,}

\def\etal{et al.$\;$}


\def\kms{km\thinspace s$^{-1}$}
\def\Lsun{L$_\odot$}
\def\Msun{M$_\odot$}
\def\ms{m\thinspace s$^{-1}$}
\def\percc{cm$^{-3}$}
\title{Distribution of water in the G327.3--0.6 massive star-forming region\thanks{{\it Herschel} is an ESA space observatory with science instruments provided by European-led Principal Investigator consortia and with important participation from NASA.}}

\titlerunning{The distribution of water in the G327.3--0.6 massive star-forming region}

\subtitle{}

\author{S. Leurini
        \inst{1,2}
        \and
        F. Herpin
        \inst{3}
        \and
        F. van der Tak
        \inst{4,5}
        \and
        F. Wyrowski
        \inst{1}
        \and
        G.J. Herczeg
        \inst{6}
        \and
        E. F. van Dishoeck
        \inst{7,8}
}

\institute{
Max-Planck-Institut f\"ur Radioastronomie, Auf dem H\"ugel 69, 53121 Bonn, Germany,
        \email{sleurini@mpifr-bonn.mpg.de}
        \and
        INAF-Osservatorio Astronomico di Cagliari, Via della Scienza 5, I-09047, Selargius (CA)
        \and
Laboratoire d'astrophysique de Bordeaux, Univ. Bordeaux, CNRS, B18N, all\'ee Geoffroy Saint-Hilaire, 33615 Pessac, France
\and 
SRON Netherlands Institute for Space Research, PO Box 800, 9700AV, Groningen, The Netherlands
\and
Kapteyn Astronomical Institute, University of Groningen, The Netherlands
\and
Kavli Institut for Astronomy and Astrophysics, Yi He Yuan Lu 5, HaiDian Qu, Peking University, Beijing, 100871, PR China
\and  
Leiden Observatory, Leiden University, PO Box 9513, 2300 RA, Leiden, The Netherlands
\and 
Max-Planck-Institut f\"ur Extraterrestrische Physik, Giessenbachstrasse 1, 85748 Garching, Germany
}

\date{\today}

\abstract
   {}  
   {Following our past study of the distribution of warm gas in the G327.3--0.6 massive star-forming region, we aim here at characterizing the large-scale distribution of water in this active region of massive star formation made of individual objects in different evolutionary phases. We investigate possible variations of the water abundance as a function of evolution.}
   {We present Herschel/PACS (4$\arcmin \times 4\arcmin$) continuum maps at 89 and179 $\mu$m encompassing the whole region (HII region and the infrared dark cloud, IRDC) and an APEX$/$SABOCA (2$\arcmin \times 2\arcmin$) map at 350 $\mu$m of the IRDC. New spectral Herschel$/$HIFI maps toward the IRDC region covering the low-energy water lines at 987 and 1113 GHz (and their H$_2$$^{18}$O counterparts) are also presented and combined with HIFI pointed observations 
 toward the G327 hot core region. We infer the physical properties of the gas through optical depth analysis and radiative transfer modeling of the HIFI lines.}
   {The distribution of the continuum emission at 89 and 179 $\mu$m follows the thermal continuum emission observed at longer 
wavelengths, with a peak at the position of the hot core and a secondary peak in the H{\sc ii} region, and an arch-like layer of hot gas west of this H{\sc ii} region. The same morphology is observed in the  \hoK~line, in absorption toward all submillimeter dust condensations. Optical depths of approximately 80 and 15 are estimated and correspond to column densities of $10^{15}$ and $2\times10^{14}$ cm$^{-2}$ , respectively, for the hot core and IRDC position. These values indicate an abundance of water relative to H$_2$ of 3$\times 10^{-8}$ toward the hot core, while the abundance of water does not  change along the IRDC with values close to some 10$^{-8}$. Infall (over at least 20$\arcsec$) is detected toward the hot core position with a rate of $1-1.3\times 10^{-2}$ \Msol$/$yr, high enough to overcome the radiation pressure that is due to the stellar luminosity. The source structure of the hot core region appears complex, with a cold outer gas envelope in expansion, situated between the outflow and the observer, extending  over 0.32 pc. 
 The outflow is seen face-on and rather centered away from the hot core.}  
   {The distribution of water along the IRDC is roughly constant with an abundance peak in the more evolved object, that is, in the hot core. These water abundances are in agreement with previous studies in other massive objects and chemical models.}
   \titlerunning{Water in G327.3--0.6}
   \maketitle
%

\section{Introduction}

In the past years, several studies have focused on the characterization of
water, a crucial molecule in modeling the chemistry and the physics of
molecular clouds \citep{dishoeck2014p}, in
different environments of star formation. In particular, the
key program Water In
Star-forming regions with Herschel (WISH)
\citep{vandishoeck2011} targeted different phases of star and
planet formation to understand the evolution of water in these
sources, while other {\it Herschel} projects also investigated water
in selected sources \citep[e.g.,][]{2013ApJ...765...61E,2014A&A...568A.125S,2014A&A...564L..11L,2015ApJ...799..102G}. 
Most of these studies focus on observations of single sources and do not contain much spatial information
on the distribution of water in the environment surrounding the source. Exceptions are the work of \citet{jacq2016}, which covered the clouds surrounding the mini-starburst W43 MM1, or studies of
large-scale molecular outflows \citep[e.g.,][]{2013A&A...549A..16N,2014A&A...568A.125S}, but in these cases,
only the immediate surrounding of low-mass young stellar objects was investigated.

\begin{table*}
\caption{\textit{Herschel}/HIFI observed water line transitions (toward the hot core region in pointing mode). Frequencies are from \citet{pearson1991}. The rms is the noise in $\delta \nu=1.1$MHz.}     
\label{table_transitions}    
\centering                   
\begin{tabular}{lccccccccc}  
\hline\hline                 
Water species & Frequency &  Wavelength & $E_u$    & HIFI & Beam  &  $\eta_{\textrm{mb}}$  &  $T_{\textrm{sys}}$   &  rms & obsid \\   
                      &    [GHz]     &       [$\mu$m] &      [K]   &band &  [\arcsec]  &  & [K]  & [mK] & \\
\hline                        
   \hoA$^a$    & 547.6764     &  547.4 &  60.5    & 1a & 37.8  & 0.62 & 80 & 78 & 1342205525 \\
   \hoB    & 552.0209    &  543.1  & 61.0     &  1a & 37.8 &  0.62 & 70  & 40 & 1342191554-5\\     
   \hoF     & 994.6751  &  301.4  &  100.6   & 4a & 21.1   &  0.63 & 290  & 44  & 1342203171\\
   \hoG    & 1095.6274  &  273.8  &  248.7   &  4b & 19.9  &  0.63 & 380   & 59  & 1342214424\\
   \hoI    & 1101.6982  &  272.1  &   52.9   &  4b & 19.9 &  0.63 &  390   & 38  & 1342214422-3,1342214425-6\\
   \hoJ    & 1107.1669  &  272.1  &   52.9   &   4b & 19.9 &  0.63 &  380   & 59 & 1342214424\\
   \hoM    & 1662.4644  &  180.3  &  113.6   & 6b  & 12.7 &  0.58 & 1410   & 232  & 1342192585\\
\hline
   \hoC$^a$    & 556.9361    &  538.3  &   61.0   & 1a & 37.1  & 0.62 & 80   & 78 & 1342205525 \\
   \hoD    & 752.0332    &  398.6  &  136.9   & 2b & 28.0   &  0.64 &  90   & 50 & 1342205844\\
      \hoP    & 970.3150    &   309.0 &  598.8   & 4a & 21.8   &  0.63 &  620   & 40 & 1342227539\\
  \hoE$^a$     & 987.9268 &  303.5  &   100.8  & 4a & 21.3  & 0.63 &  340   & 65 & 1342203169-1342203170\\
   \hoH    & 1097.3651  &  273.2  &  249.4   & 4b  & 19.9  &  0.63 & 380   & 59  & 1342214424\\
   \hoK$^b$    & 1113.3430  &  269.0  &  53.4   &4b & 19.7   &  0.63 & 395   & 38 & 1342214421-3,1342214425-6\\
   \hoL    & 1661.0076  &  180.5  &  194.1   &6b & 12.7  &  0.58 & 1410   & 232 &1342192585\\
   \hoN    & 1669.9048  &  179.5  &   114.4   & 6b  & 12.6  &  0.58 & 1410   & 232 & 1342192585\\
\hline                                  
\end{tabular}
\tablefoot{$^a$ This line was mapped in OTF mode (small map). $^b$ This line was mapped in OTF mode (large map).}
\end{table*}

As part of the WISH project, 
six nearby cluster-forming clouds were mapped in multiple water transitions; the data were
complemented with mid- and high-J CO and $^{13}$CO observations with the APEX telescope and {\it Herschel} to better
characterize the warm gas in the (proto-) clusters.  
In this paper, we present   observations of water of the star-forming region G327.3--0.6  
 at a distance of 3.1\,kpc \citep{2015A&A...579A..91W}.
 Different evolutionary phases of massive star formation coexist in a small ($\sim$3pc) region \citep{2006A&A...454L..91W}:
 a bright H{\sc ii} region \citep{1970AuJPA..14....1G} associated with a luminous photon-dominated region seen in CO
 \citep[hereafter Paper\,I,][]{2013A&A...550A..10L}, and a chemically extremely rich hot core \citep{1998A&A...337..275N,2000ApJ...545..309G} in a
 cold infrared dark cloud hosting several other dust condensations \citep{2009A&A...501L...1M}, one of which
has signs of active star formation \citep{2008AJ....136.2391C}. The region was studied in mid-J CO
and $^{13}$CO lines in Paper\,I:  emission is detected over the whole extent of the maps ($3\times4$ pc)
with excitation temperatures ranging from 20\,K up to 80\,K in the gas around the H{\sc ii} region,
and H$_2$ column densities from a few $10^{21}$~cm$^{-2}$  in the 
interclump gas to $3\times 10^{22}$~cm$^{-2}$ toward the hot core.
The warm gas is only a small percentage ($\sim$10\%) of the total gas in the infrared dark cloud,
while it reaches values of up to 
$\sim$35\% of the total gas in the ring surrounding the H{\sc ii} region.
The goal of our current study is to
characterize the large-scale distribution of water in an active region of massive star formation that shows different
evolutionary phases to verify whether its abundances varies as a function of evolution. 

\section{Observations and data reduction}

We present mapping observations of the G327.3--0.6 massive star-forming
region collected with the HIFI \citep{2010A&A...518L...6D} and PACS \citep[][]{poglitsch2010} spectroscopic instruments on board {\it Herschel}\footnote{Data can be retrieved from the {\it Herschel} Archive System, \url{http://archives.esac.esa.int/hsa/whsa}} \citep{2010A&A...518L...1P} in the framework of the WISH program. Additional APEX observations with the SABOCA camera (Sect.~\ref{apex}) are also discussed. 

\subsection{HIFI pointed observations and maps}
\label{hifi_obs}

Three water lines as well as the $^{13}$CO(10--9) \citep[][]{2013A&A...550A..10L} and C$^{18}$O(9--8) lines have been observed with HIFI in August 2010 (OD 461) and February 2011 (OD 645) using the on-the-fly observing mode with Nyquist sampling. The center of the map is $\alpha_{J2000}$= 15h53m05.48s, $\delta_{J2000}$ =-54$^{\circ}$36'06.2". The reference position was 5 arcmin offset north in declination for all observations.

The HIFI observations were made in bands 4B and 4A. The sideband separation of 8 GHz and IF bandwidth of 4 GHz allow a local oscillator (LO) setting where the o-H$_2$O and H$_2$$^{18}$O $1_{11}-0_{00}$ transitions at 1113.343 GHz and 1101.698 GHz, respectively, and the $^{13}$CO(10--9)  transition at 1101.350 GHz can be observed simultaneously. The same holds for the p-H$_2$O and C$^{18}$O (9--8) transitions at 987.927 GHz and 987.560 GHz, respectively. The 1113\,GHz water map consists of 19 OTF rows made of 26 independent points covering $3.5' \times 2.7'$ (one map coverage), while the 987 water map consists of 8 OTF rows made of 10 independent points covering $1.2' \times 1.2'$ (with two map coverages).

As with all massive protostars observed by the WISH GT-KP, 14 water lines (see~Table \ref{table_transitions}) were observed with HIFI in the pointed mode at frequencies between 547 and 1670 GHz in 2010 and 2011 (list of observation identification numbers, {\em obsids}, are given in Table \ref{table_transitions}) toward the G327 hot core region (RA=15h53m08.8s, DEC= -54$^{\circ}$37'01"), between SMM2 and the hot core position (see Sect. \ref{sec:HIFIanalysis}) because of a confusion between different references \citep[e.g.,][]{1992PhDT.......252B}. An additional high-energy water line at 970.3150 GHz was also observed. We used the double beam-switch observing mode with a throw of 3'. The off positions were inspected and did not show any emission. The frequencies, energy of the upper levels, system temperatures, integration times, and {\it rms} noise level at a given spectral resolution for each of the lines are provided in Table \ref{table_transitions}. 

Data were taken simultaneously in H and V polarizations using both the acousto-optical Wide-Band Spectrometer (WBS) with 1.1 MHz resolution and the digital auto-correlator or High-Resolution Spectrometer (HRS), which provides higher spectral resolution. Calibration of the raw data into the $T_A$ scale was performed by the in-orbit system \citep[][]{roelfsema2012}; conversion to $T_{mb}$ was made using the latest beam efficiency estimate from October 2014\footnote{http://www.cosmos.esa.int/web/herschel/home} given in Table~\ref{table_transitions} and a forward efficiency of 0.96. HIFI receivers are double sideband with a sideband ratio close to unity \citep[][]{roelfsema2012}. The flux scale accuracy is estimated to be between 10\% for bands 1 and 2, 15\% for bands 3 and 4, and 20\% in bands 6 and 7$^1$. The frequency calibration accuracy is 20 kHz and 100 kHz (i.e., better than 0.06 \kms) for HRS and WBS observations, respectively. Data calibration was performed in the Herschel Interactive Processing Environment \citep[HIPE,][]{ott2010} version 12. Further analysis was made within the CLASS\footnote{http://www.iram.fr/IRAMFR/GILDAS/} package (Dec. 2015 version). These lines are not expected to be polarized, therefore data from the two polarizations were averaged together after inspection. For all observations, eventual contamination from lines in the image sideband of the receiver was checked and none was found. Some unidentified features (not due to water species) are nevertheless detected but not blended with the water lines. Because HIFI is operating in double-sideband, the measured continuum level was divided by a factor of 2 (in the figures and tables) to be directly compared to the single-sideband line profiles (this is justified because the sideband gain ratio is close to 1). 

\begin{table}
\centering
\caption{Summary of the  PACS {\it Herschel}  and SABOCA APEX observations.}
\label{PACS_sum}
\begin{tabular}{lrcccccc}
\hline
\hline
Transition &$\lambda$&$E_u$&&$$R$$&r.m.s.\\
&[$\mu$m]&[K]&[\arcsec]& &Jy$/$beam\\
\hline
\multicolumn{6}{c}{PACS}\\
$o$-H$_2$O $2_{12}-1_{01}$&179.53&114.4&12.3&1467& 1\\
$o$-H$_2$O $3_{03}-2_{12}$&174.63&196.8&12.0&1409&1 \\
\hline
\multicolumn{6}{c}{Continuum observations}\\
SABOCA &350&&7.8& -- & 2\\
PACS  & 89.8&&  9.1 & -- & 2 \\
PACS  &179.5& &12.3& -- & 4 \\
\hline
\hline
\end{tabular}
\end{table}

\subsection{PACS maps}\label{pacs_obs}

\begin{figure}
\centering
\resizebox{\hsize}{!}{\includegraphics{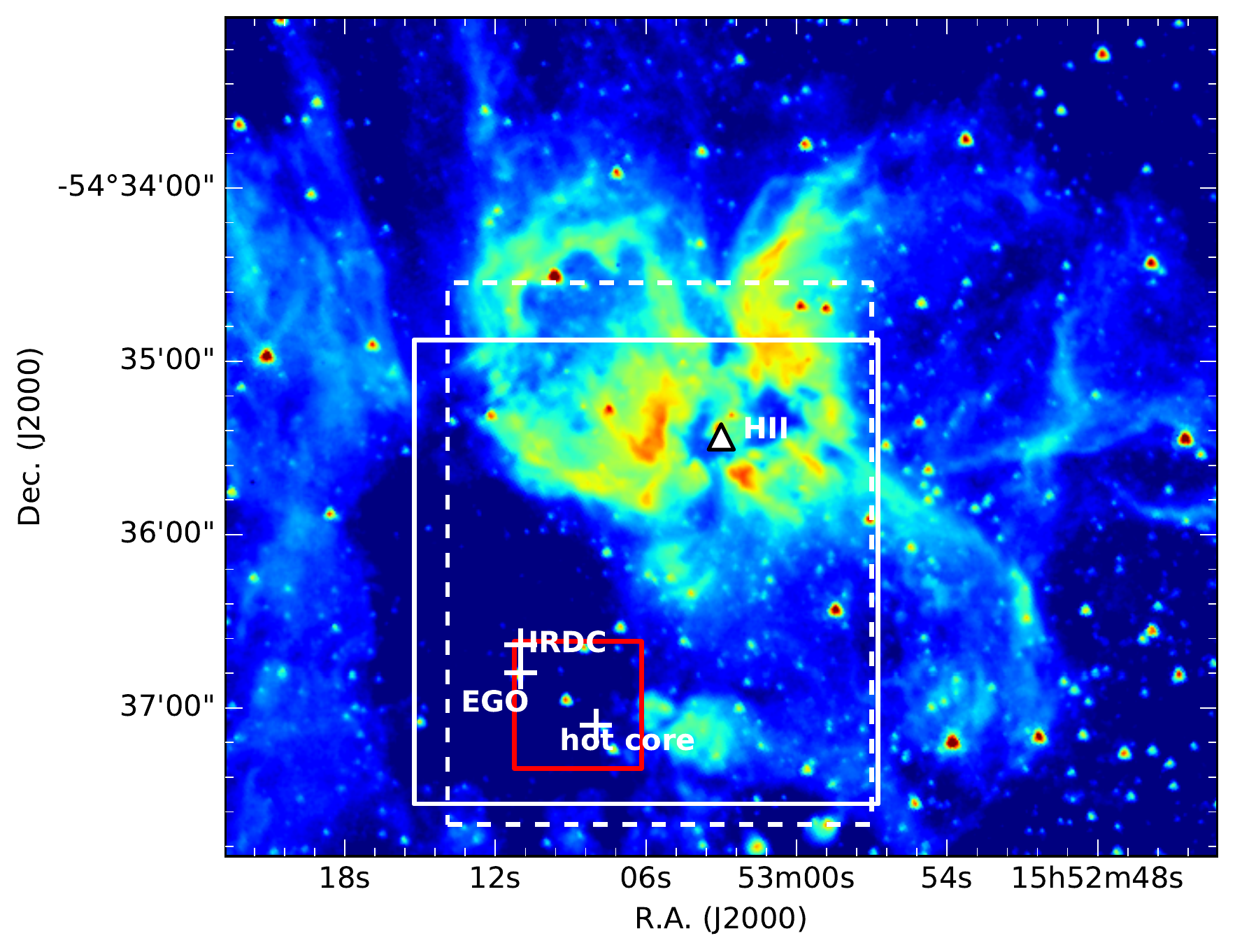}}
\caption{Large-scale {\it Spitzer} image at 3.6 $\mu$m of G327.36--0.6. The boxes show the areas mapped with PACS at 89 and 179 $\mu$m (solid white line), with HIFI at 1113\,GHz (white dashed line), and at 987\,GHz (red line). The white crosses and triangle mark the positions discussed in Paper~I.}\label{i1}
\end{figure}

PACS is an integral field unit with a $5 \times 5$ array of spatial pixels (hereafter spaxels). Each spaxel covers $9\farcs4 \times 9\farcs4$, providing a total field of view of $\sim 47\arcsec\times47\arcsec$. The observations (see Table \ref{PACS_sum}, obsid 1342192145) were performed using the PACS chopped line spectroscopic mode \citep[see][]{poglitsch2010}. The area mapped with PACS is shown in Fig. \ref{i1}. This mode achieves a spectral resolution of $\sim0.12$ $\mu$m (corresponding to a velocity resolution of $\sim 210$ km$/$s). 
Two nod positions were used that chopped 6' on each side of the source. The two positions were compared to assess the influence of the off-source flux of observed species from the off-source positions. The typical pointing accuracy is better than 2\arcsec. 

We performed the basic data reduction with the Herschel Interactive Processing Environment v.12 (HIPE, Ott 2010). The flux was normalized to the telescope background and calibrated using Neptune observations. Spectral flatfielding within HIPE was used to increase the signal-to-noise
ratio \citep[for details, see ][]{2012A&A...540A..84H, green2016}. In order to account for the substantial flux leakage between the spaxels surrounding the true source position and to improve the continuum stability, custom IDL routines were used to further process the datacubes for the wavelength-dependent loss of radiation for a point source (see PACS Observers Manual). The overall flux calibration is accurate to $\sim20 \%$ based on the flux repeatability for multiple observations of the same target in different programs, cross-calibrations with HIFI and ISO, and continuum photometry. The continuum (and line) rms are given in Table \ref{PACS_sum}.  

\subsection{SABOCA map}\label{apex}
The IRDC in G327.36--0.6 was observed with the APEX\footnote{APEX is a collaboration between
the Max-Planck-Institut f\"ur Radioastronomie, the European Southern
Observatory, and the Onsala Space Observatory.} telescope in the continuum emission at 350 $\mu$m with 
the Submillimeter APEX Bolometer Camera \citep[SABOCA, ][]{2010Msngr.139...20S}. The observations were performed in 2010,
on May 11 (see Table \ref{PACS_sum}). The pointing was checked on B13134 (also used as flux calibrator) and on the bright hot core hosted in the IRDC, on the peak of the 3~mm continuum emission
obtained by \citet{2008Ap&SS.313...69W} with the ATCA array. Skydips (fast scans in elevation at constant azimuthal angle) were performed to estimate the atmospheric opacity. The weather conditions at the time of the observations were good, with a median precipitable water vapor level of 0.24~mm. The data were reduced with the BOA software \citep{2012SPIE.8452E..1TS}.

\section{Observational results}
 
\subsection{Continuum emission}
\label{cont_emi}

\begin{figure}
\centering
\resizebox{\hsize}{!}{\subfigure[][]{\includegraphics{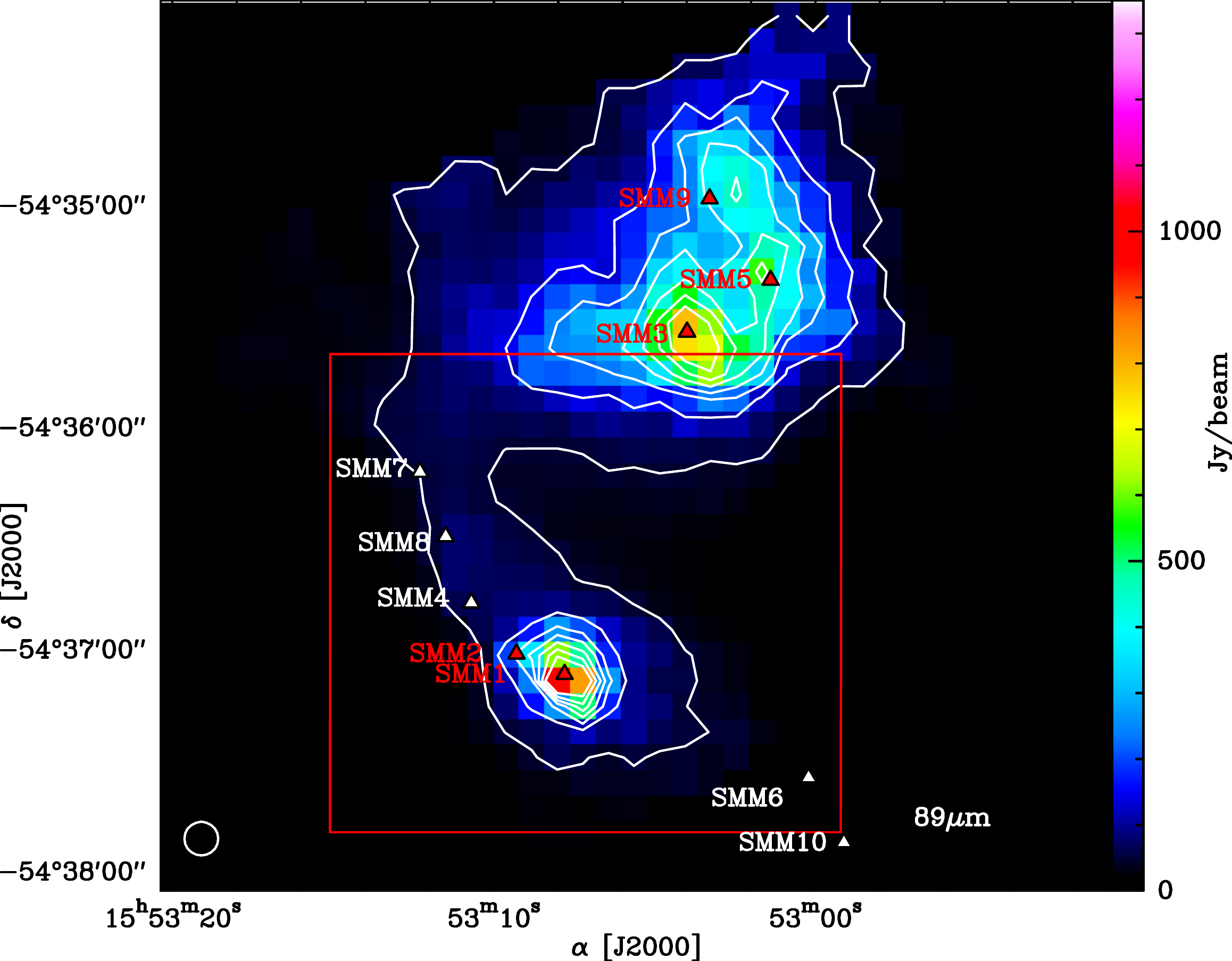}}}
\resizebox{\hsize}{!}{\subfigure[][]{\includegraphics{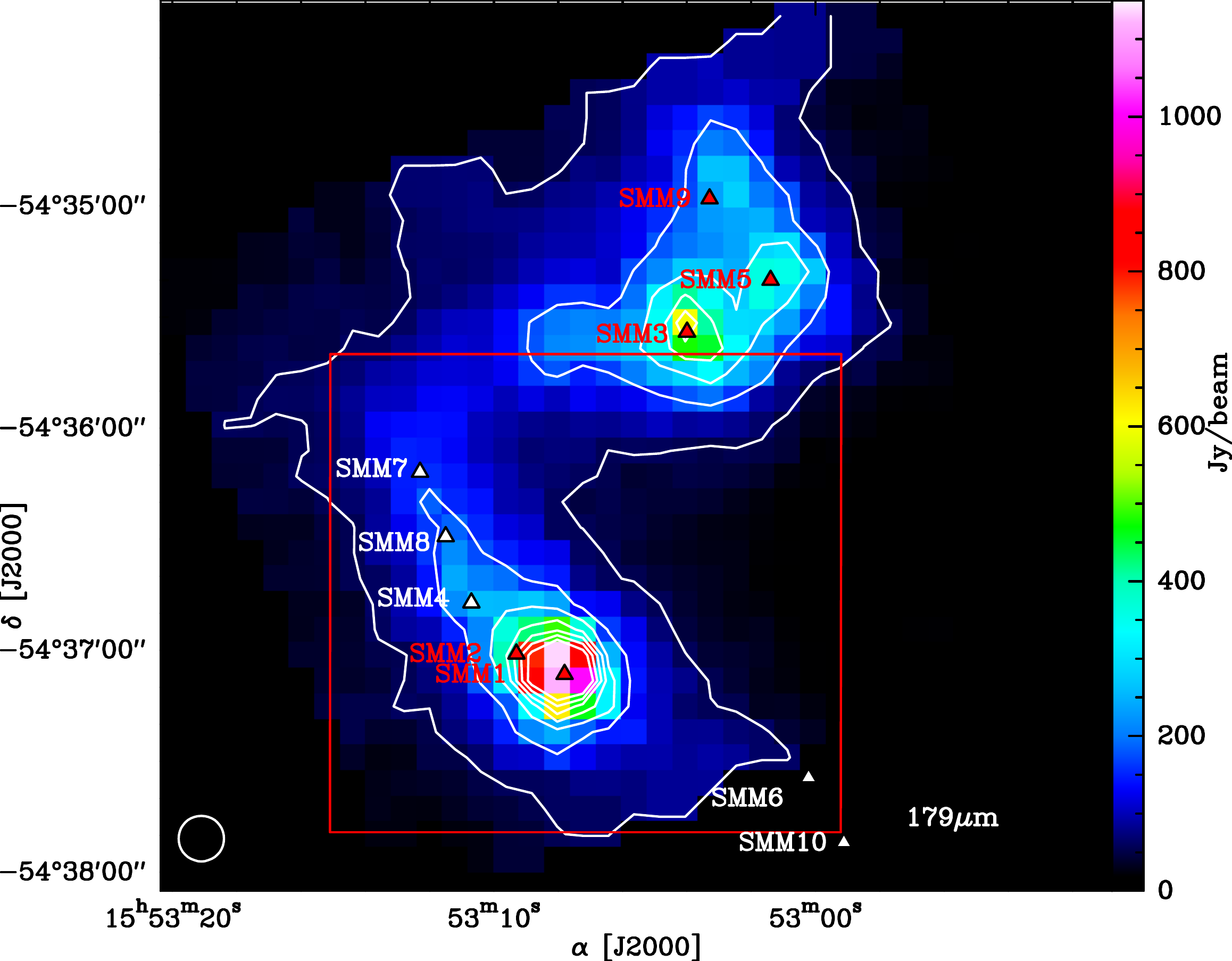}}}
\caption{Color scale and white contours  are the PACS continuum image of G327.36--0.6 at 89 (top panel, resolution is 9.1$\arcsec$)  and 179 $\mu$m (bottom panel, resolution is 12.3$\arcsec$). Contours are  from 5\% of the peak flux in steps of 10\%. The triangles mark the positions of the submillimeter continuum peaks reported in Table~\ref{pos}. The red box outlines the area plotted in Fig.~\ref{saboca}.}
\label{pacs}
\end{figure}

\begin{table}
\centering
\caption{Overview of the sources in the G327.3--0.6 massive star-forming region (positions corrected by the shift as explained in Sect. \ref{cont_emi}).}\label{pos}
\begin{tabular}{lcc}
\hline
\hline
\multicolumn{1}{c}{source} &\multicolumn{1}{c}{$\alpha_{\rm{J2000}}$}&\multicolumn{1}{c}{$\delta_{\rm{J2000}}$}\\
\hline
SMM1 (hot core)$^{a,b}$&15h53m07.8s&-54$^{\circ}$37'06.5"\\
SMM2$^{\rm{a}}$&15h53m09.3s&-54$^{\circ}$37'01.0"\\
SMM3$^{\rm{c}}$&15h53m04.0s&-54$^{\circ}$35'34.0"\\
SMM4$^{\rm{a}}$&15h53m10.7s&-54$^{\circ}$36'47.2"\\
SMM5$^{\rm{c}}$&15h53m01.4s&-54$^{\circ}$35'20.0"\\
SMM6$^{\rm{a}}$&15h53m00.2s&-54$^{\circ}$37'34.4"\\
SMM7$^{\rm{a}}$&15h53m12.3s&-54$^{\circ}$36'12.9"\\
SMM8$^{\rm{a}}$&15h53m12.1s&-54$^{\circ}$36'31.0"\\
SMM9$^{\rm{a}}$&15h53m03.3s&-54$^{\circ}$34'58.0"\\
SMM10$^{\rm{a}}$&15h52m59.1s&-54$^{\circ}$37'52.0"\\
EGO$^{\rm{d}}$&15h53m11.2s&-54$^{\circ}$36'48.0"\\
H{\sc ii}$^{\rm{e}}$&15h53m03.0s&-54$^{\circ}$35'25.6"\\
\hline

\end{tabular}
\begin{list}{}{}
\item[$^{\rm{a}}$] based on the SABOCA map
\item[$^{\rm{b}}$] the ATCA 3~mm position of \citet{2008Ap&SS.313...69W} is $\alpha_{\rm{J2000}}=15^h53^m07^s.8, \delta_{\rm{J2000}}=-54\degr37\arcmin06\farcs4$.
\item[$^{\rm{c}}$] \citet{2009A&A...501L...1M}
\item[$^{\rm{d}}$] \citet{2008AJ....136.2391C}
\item[$^{\rm{e}}$] peak of the centimeter continuum 
emission from ATCA archival data at 2.3~GHz, project number C772
\end{list}
\end{table}

\begin{figure}
\centering
\resizebox{\hsize}{!}{\includegraphics{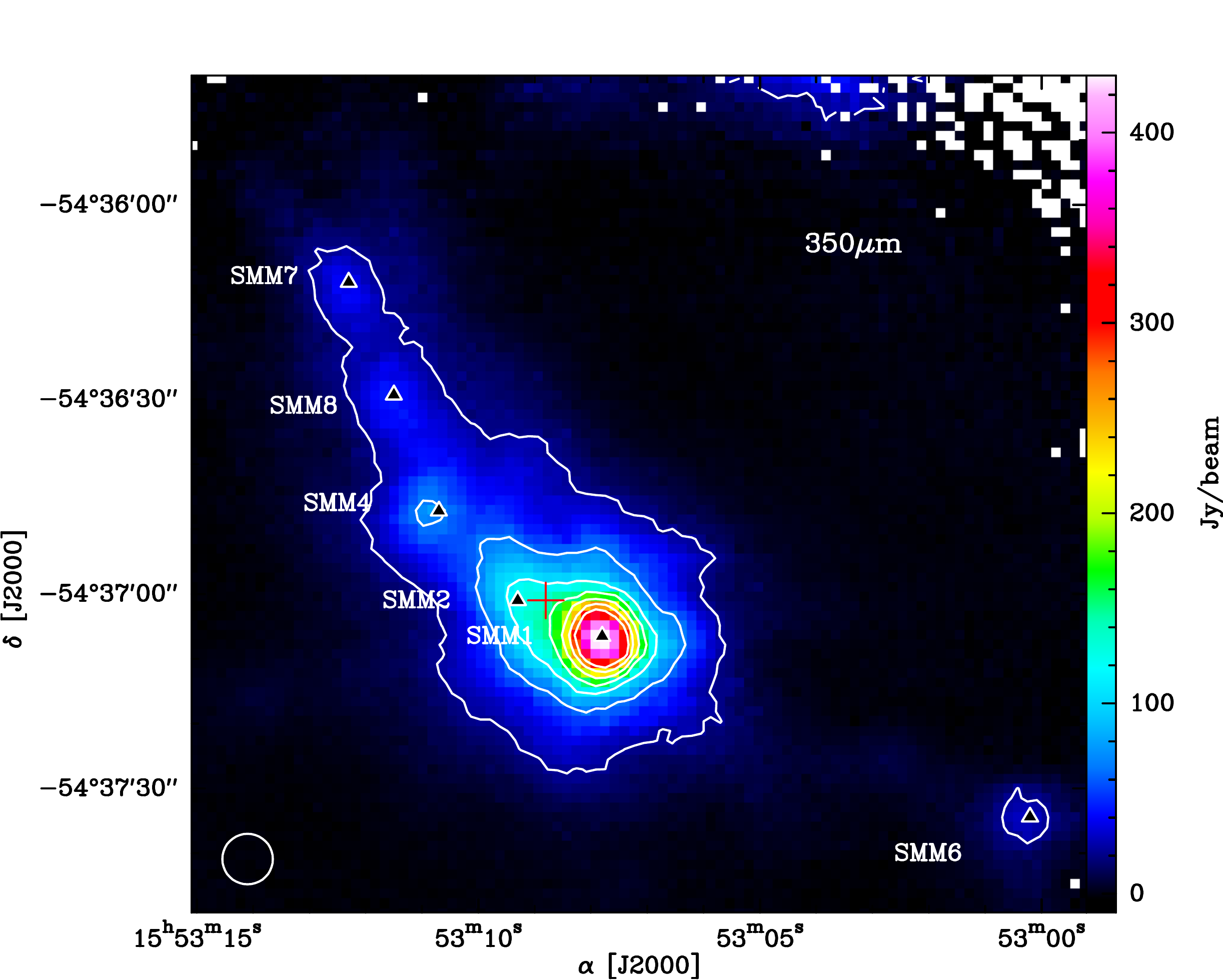}}
\caption{Distribution of the  SABOCA continuum emission at 350 $\mu$m  along the infrared dark cloud in G327.3--0.6. Contours are from 5\% of the peak flux in steps of 10\%. The triangles mark the positions of the submillimeter continuum peaks reported in Table~\ref{pos}. The red cross marks the position observed for the single pointing HIFI observations. The white circle in the bottom left corner shows the beam of the SABOCA observations.}
\label{saboca}
\end{figure}

Figure~\ref{pacs} shows the distribution of the continuum emission of G327.3--0.6 at 89 and 179 $\mu$m 
observed with PACS. The morphology follows the thermal continuum emission observed at larger 
wavelengths \citep{2009A&A...504..415S,2009A&A...501L...1M}, with a peak at the position of the hot core
and a secondary peak in the H{\sc ii} region toward SMM3. Additionally, the 179 $\mu$m map also shows weak emission along an arch-like layer of hot gas west of the H{\sc ii} region seen in Fig.~\ref{i1} at 3.6 $\mu$m but also in $^{12}$CO and $^{13}$CO (Paper~I).
The SABOCA map of the IRDC in the G327.3--0.6 massive star-forming
region is shown in Fig.~\ref{saboca}. The map shows a shift toward
the east with respect to the continuum map at 450 $\mu$m published by
\citet{2009A&A...501L...1M}. However, the peak of the
350 $\mu$m continuum emission coincides with the position derived for
the hot core in Paper~I and with the position inferred with
interferometric measurements at 3~mm \citep{2008Ap&SS.313...69W}
within $\sim$1\farcs3, while the peak of the 450 $\mu$m continuum map
is shifted of (7\arcsec,-2.6\arcsec) from the ATCA position. Therefore the difference between the two continuum maps is probably
due
to a pointing error in the 450 $\mu$m data (larger than their pointing accuracy), which then have been shifted.

We used the Gauss-clump
program \citep{1990ApJ...356..513S,1998A&A...329..249K} to derive the
positions of the dust condensations discussed by
\citet{2009A&A...501L...1M}. Their new coordinates are reported in
Table~\ref{pos} together with other sources in the region discussed in
Paper~I and in this study. The largest offset is for SMM6, whose SABOCA position is
(-6\farcs1, -5\farcs6) from the previous reported one, although the
source is well isolated. The other sources (SMM1, SMM2, SMM4, SMM7,
and SMM8) have a shift \citep[compared to][]{2009A&A...501L...1M} between -4\farcs3 and -7\farcs0 in right ascension,
and between -1\farcs9 and 2\farcs5 in declination from the
corresponding 450 $\mu$m sources. For the region not covered by our
SABOCA map, the coordinates listed in Table~\ref{pos} are from \citet{2009A&A...501L...1M}.

\begin{figure}
\centering
\resizebox{\hsize}{!}{\includegraphics{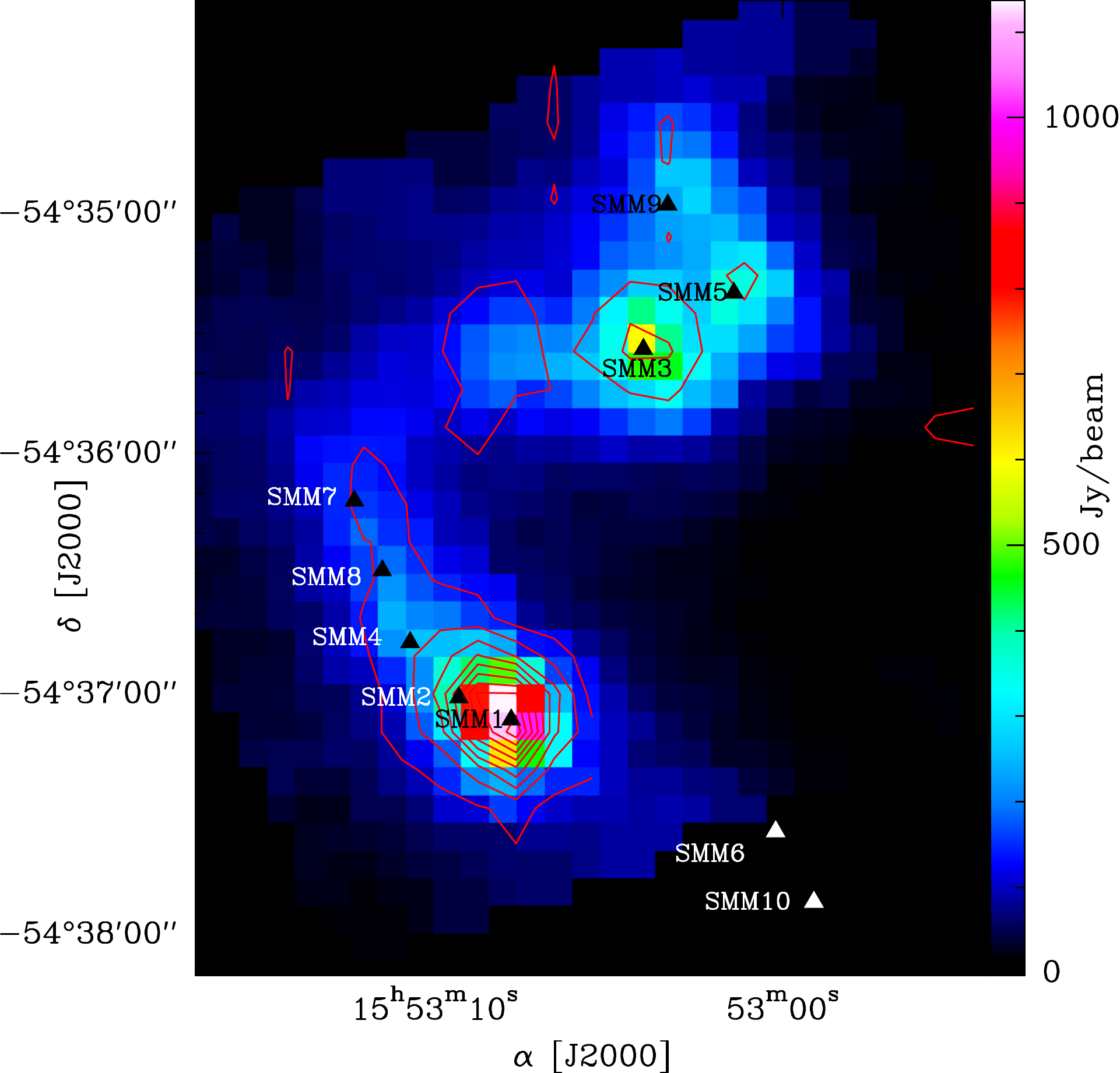}}
\caption{Distribution of the continuum emission at 179 $\mu$m  in G327.3--0.6 (color scale). The solid red 
contours represent the distribution of the absorption in the
 $1_{11}\to0_{00}$ p-H$_2$O line, integrated in the velocity range $\rm{v_{\rm{LSR}}}=[-55,-37]$~km~s$^{-1}$ 
(from -3$\sigma$,-4.5~K~km~s$^{-1}$, in steps of -3$\sigma$). Labels are the peaks of the 450 $\mu$m continuum emission from \citet{2009A&A...501L...1M}.}
\label{111}
\end{figure}

\begin{figure}
\centering
\resizebox{\hsize}{!}{\includegraphics[width=10cm]{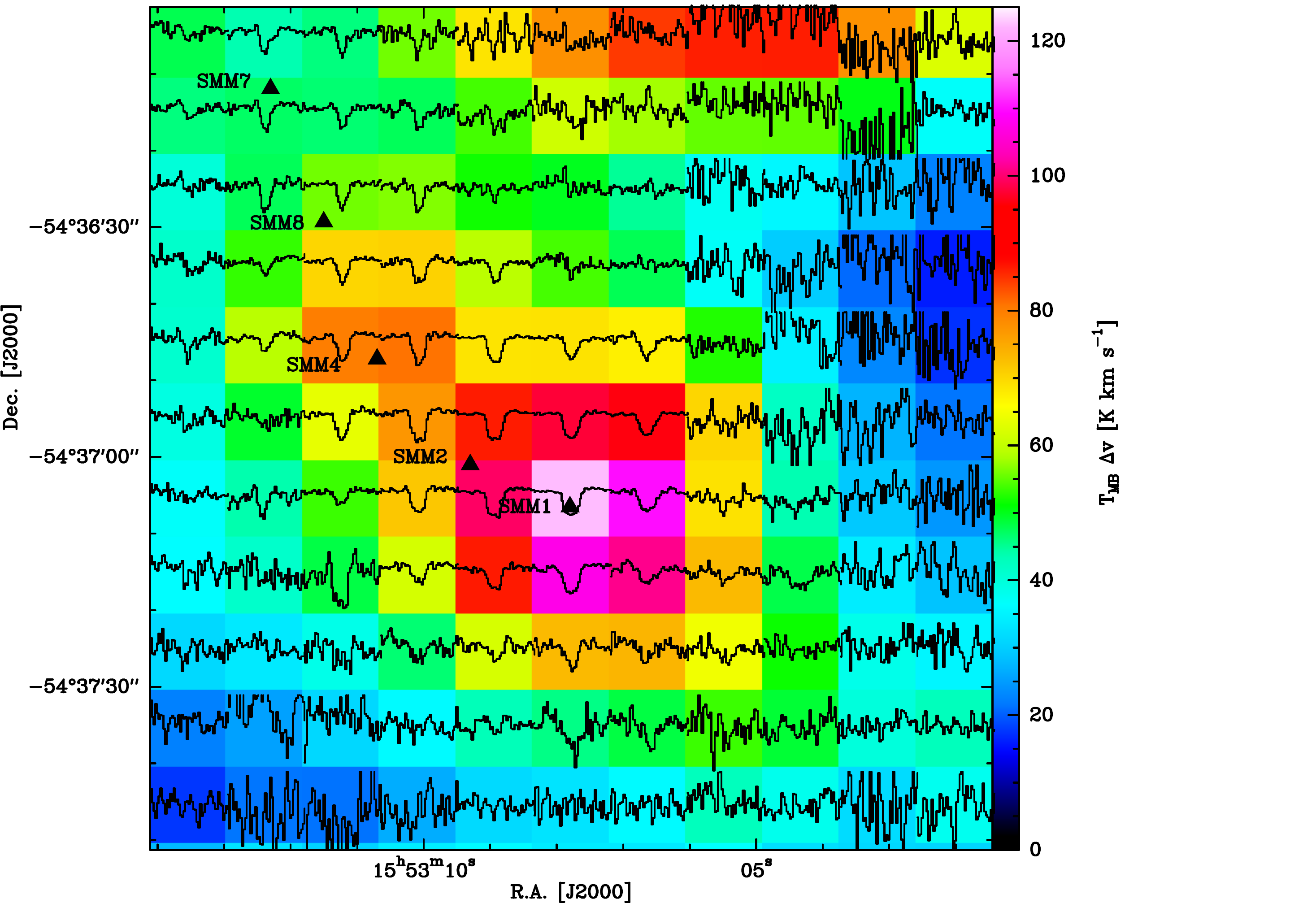}}
\caption{Spectral HIFI map of the line-to-continuum ratio of the 
$1_{11}\to0_{00}$ p-H$_2$O line toward the IRDC region overlaid on the $^{13}$CO(6--5) 
integrated emission (color image).  The temperature axis ranges from -1 to 1.5~K, the velocity axis ranges from -65 to -35~km~s$^{-1}$.
The $^{13}$CO(6--5)  data are smoothed to the resolution of the H$_2$O map. The black triangles mark the positions of the peaks of the 450\,$\mu$m continuum emission.
}\label{IRDC_1113}
\end{figure}

\subsection{Large-scale distribution of water}
\begin{figure}
\centering
\resizebox{\hsize}{!}{\includegraphics{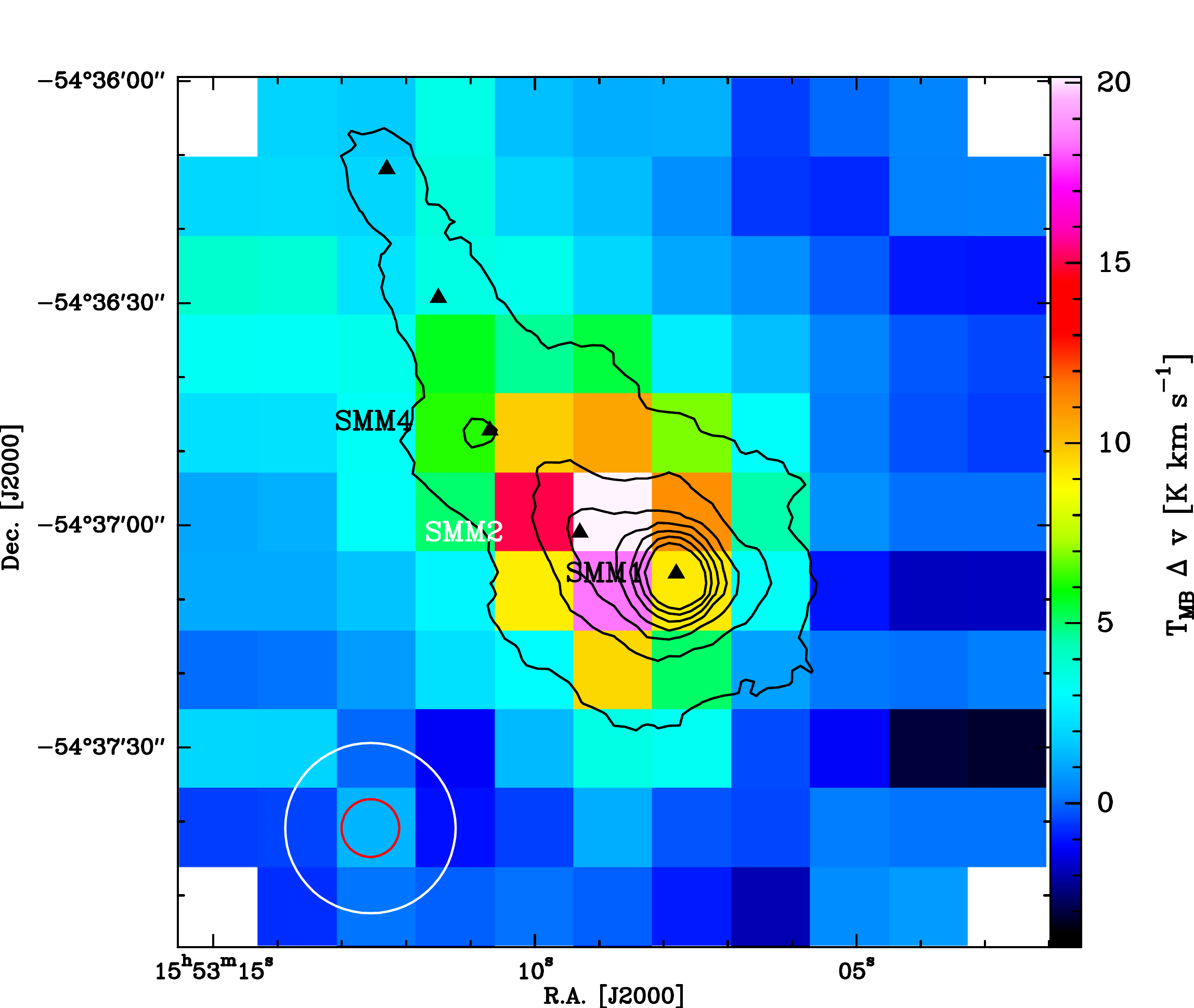}}
\caption{Integrated HIFI intensity map of the p-H$_2$O 2$_{02}$-1$_{11}$
  line ([-50,-38]\,km\,s$^{-1}$) toward the IRDC region (color image). The black contours show the
  SABOCA continuum emission at 350 $\mu$m from 5\% of the peak flux in steps of 10\%. The triangles mark the positions of the submillimeter continuum peaks reported in Table~\ref{pos}.    Beams of the observations of the p-H$_2$O 2$_{02}$-1$_{11}$
  line (white circle) and of the 350\,$\mu$m continuum  (red circle) are shown in the bottom left corner.}\label{IRDC_987}
\end{figure}

The distribution of the absorption in the \hoK~line is shown in Fig. \ref{111} and closely follows the distribution of the continuum emission at 179 $\mu$m. The detailed distribution of water in the IRDC and the HII regions are discussed in Sects. \ref{distri_IRDC} and \ref{HIIregion}, respectively.

\subsubsection{ IRDC}
\label{distri_IRDC}
 \begin{figure}
\centering
\resizebox{\hsize}{!}{\includegraphics{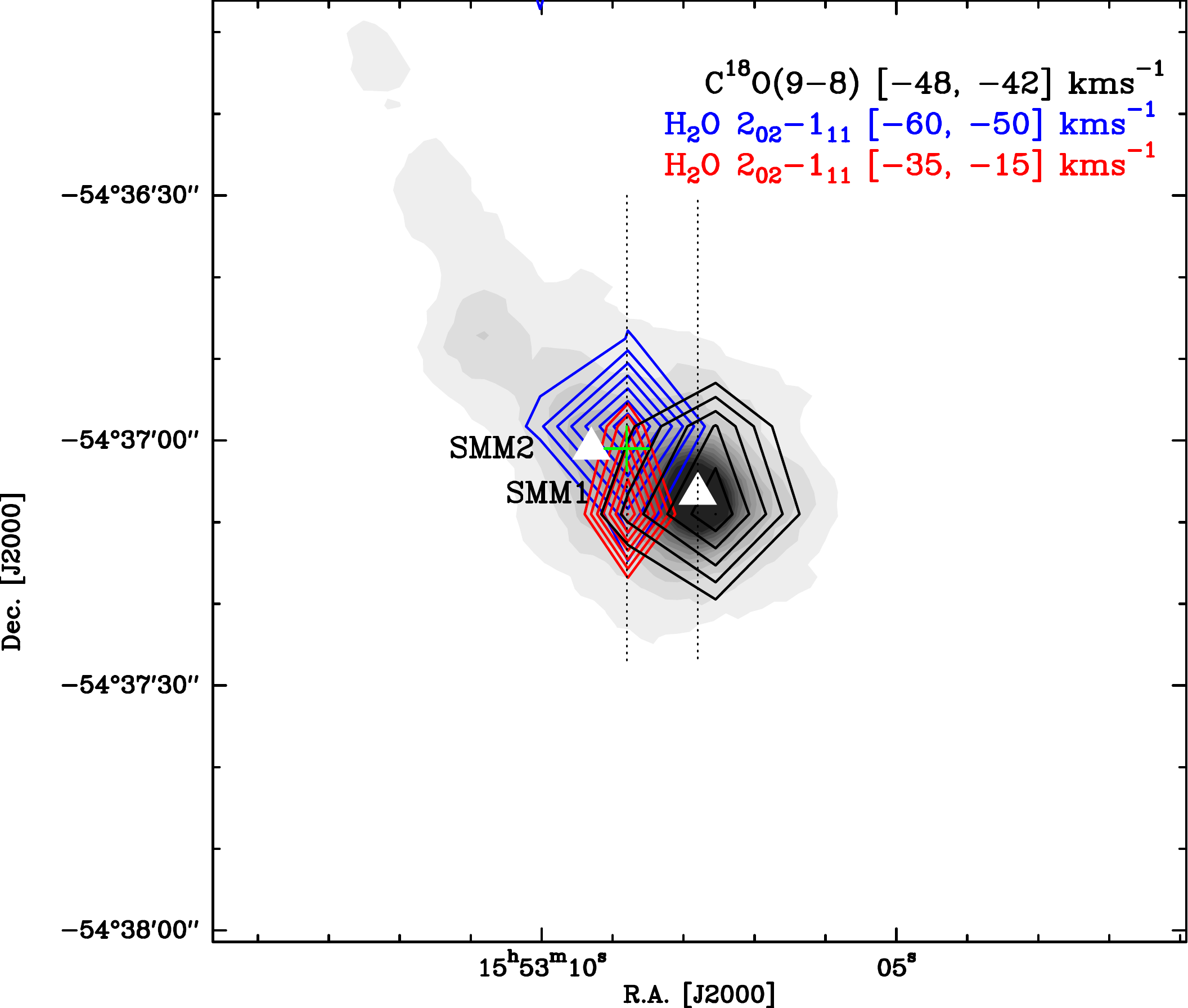}}
\caption{Integrated intensity map of the p-H$_2$O 2$_{02}$-1$_{11}$
 line in the blue- ([-60,-50]\,km\,s$^{-1}$, blue contours from 30\% of the peak emission in steps of 10\%) and redshifted
([-35,-15]\,km\,s$^{-1}$, red contours from 30\% of the peak emission in steps of 10\%) velocity ranges
toward the IRDC region. The gray contours represent the integrated intensity of C$^{18}$O(9--8)
([-48,-42]\,km\,s$^{-1}$, from 50\% of the peak emission in steps of 10\%). The white triangles mark SMM1 and SMM2, the green cross the position observed in the single-pointing HIFI observations (labeled as outflow in Fig.\,\ref{pv}.) The dotted lines outline the cuts used to derive the P-V diagrams discussed in Sect.\,\ref{distri_IRDC}).}\label{map987out}
\end{figure}

\begin{figure*}
\centering
\subfigure{\includegraphics[width=13cm]{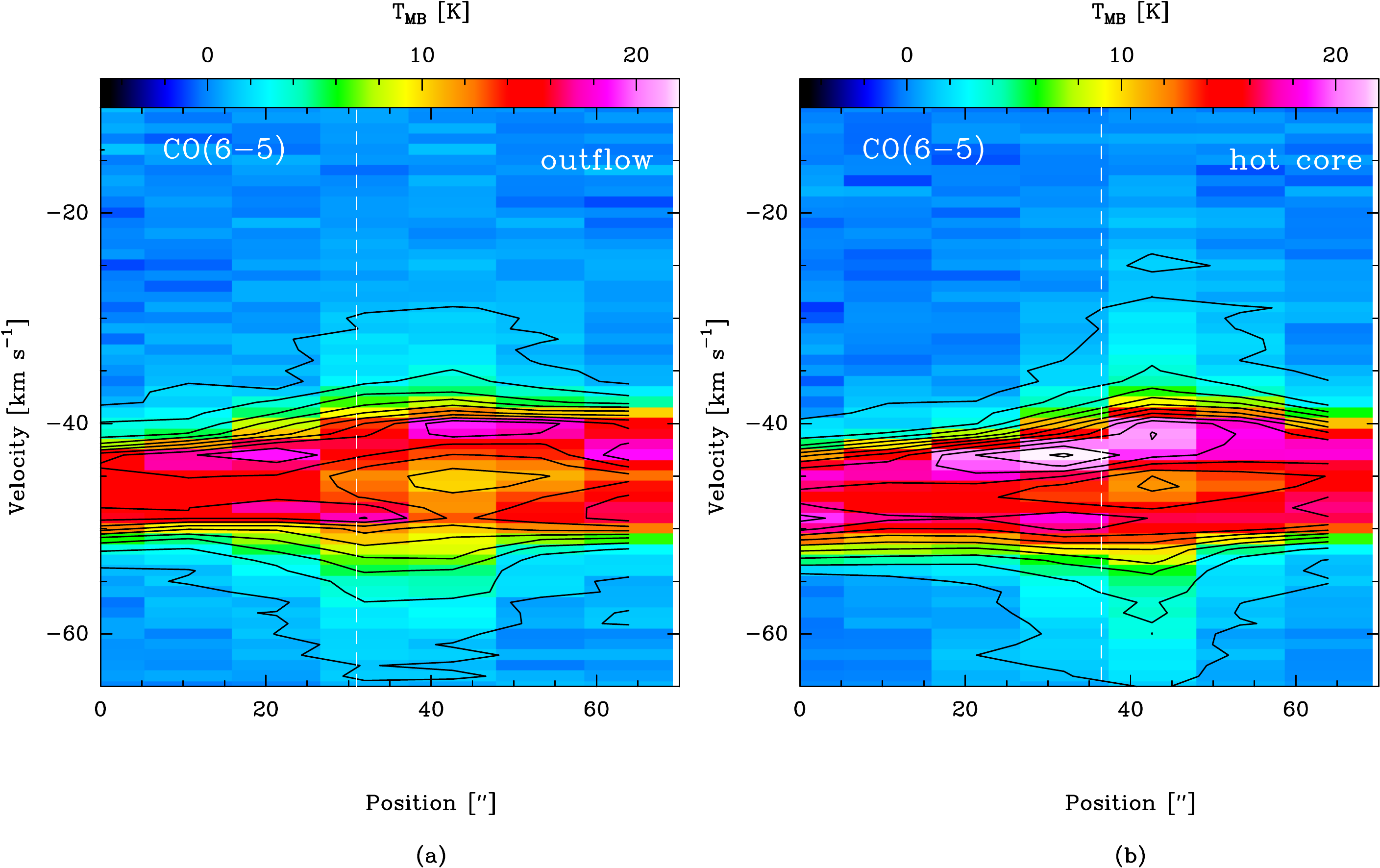}}
\subfigure{\includegraphics[width=13cm]{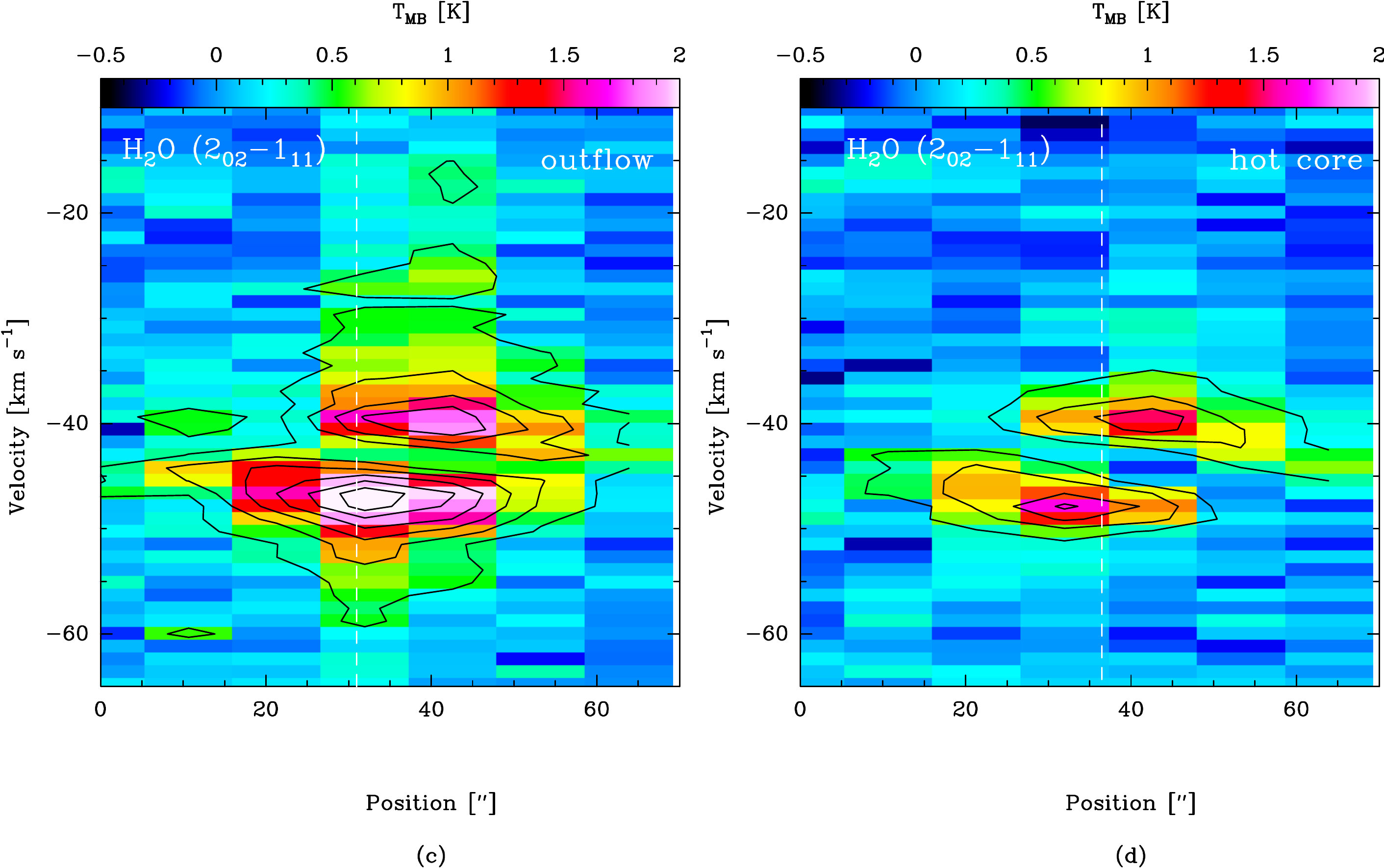}}
\caption{{\bf Top:} Color scale and contours show the P-V diagram of the CO(6--5) transition
 computed along a vertical cut passing through the outflow (a) and the hot core position (b).
{\bf Bottom:} Color scale and contours show the P-V diagram of the $2_{02}-1_{11}$ H$_2$O line  computed along a vertical cut passing through the outflow (c) and the hot core position (d).
The cut through  the outflow position  is from $\alpha_{\rm{J2000}}=15^h53^m08^s.8, \delta_{\rm{J2000}}=-54\degr36\arcmin30\arcsec$ to $\alpha_{\rm{J2000}}=15^h53^m08^s.8,  \delta_{\rm{J2000}}=-54\degr37\arcmin27\arcsec$, the cut through the hot core from $\alpha_{\rm{J2000}}=15^h53^m07^s.8, \delta_{\rm{J2000}}=-54\degr36\arcmin30\arcsec$ to $\alpha_{\rm{J2000}}=15^h53^m07^s.8, \delta_{\rm{J2000}}=-54\degr37\arcmin27\arcsec$.
Offset positions increase along the direction of the cuts. Contours are from  3$\sigma$  in steps of 3$\sigma$ for H$_2$O, and in steps of 5$\sigma$ for CO.}\label{pv}
\end{figure*}

\begin{figure}
\centering
\resizebox{\hsize}{!}{\includegraphics{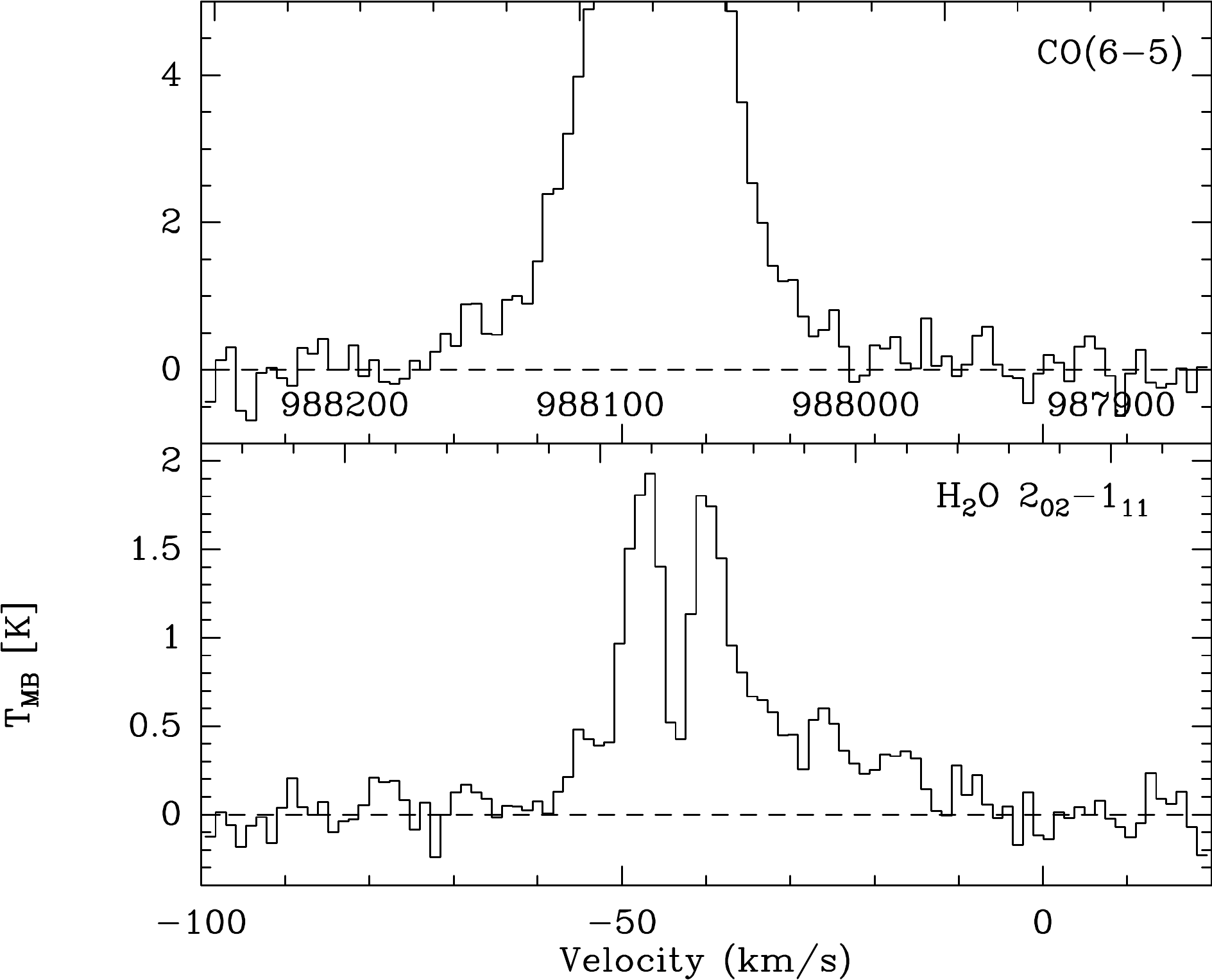}}
\caption{Spectra of the  987~GHz water line (bottom panel) and of CO(6--5)  
(top panel) at the peak of the red-shifted 
integrated intensity of the 987~GHz transition. The CO(6--5) spectrum is averaged over the 987~GHz beam.}\label{spectra-out}
\end{figure}

The IRDC hosting the hot core G327.3--0.6 was mapped in two
different transitions  of water (at 987 and 1113~GHz) with HIFI. Absorption is detected in
the 1113~GHz line toward all submillimeter dust condensations (see Fig. \ref{IRDC_1113}), but because it is saturated toward most positions, any quantitative analysis is difficult (see Sect. \ref{hifi}). The 
$2_{02}-1_{11}$ line at 987~GHz (see Fig. \ref{IRDC_987}) is seen in emission except at the positions 
of the hot core and of SMM2, where 
a combination of emission and absorption is detected. 
 The ground-state
para line shows a broad saturated absorption toward the hot core
position, and its line-width narrows along the IRDC. On the other
hand, the 987~GHz line shows broad blue- and redshifted non-Gaussian
wings. The integrated intensity maps of the red-
($\rm{v_{\rm{LSR}}}=[-35,-15]$~km~s$^{-1}$) and blueshifted
($\rm{v_{\rm{LSR}}}=[-60,-50]$~km~s$^{-1}$) 987~GHz line 
show a bipolar morphology along the north-south direction centered to the east  
of the hot core near SMM2 (see Fig. \ref{map987out}).  This shift 
is not due to a pointing error in the HIFI observations as the C$^{18}$O(9--8) line 
($\nu_{\rm{C^{18}O_{(9-8)}}}=987560.3822$~MHz, observed simultaneously to the 987~GHz water line) peaks on the hot core. The outflow is unresolved, and the blue- and 
redshifted emission is detected only in a few spectra centered approximately on (-10\arcsec,-7\arcsec)  
from the hot core. Figure~\ref{pv} shows the P-V diagrams of the CO(6--5) line (from Paper\,I, top panels) and of the 987~GHz water transition (bottom panels) along two cuts in the north-south direction passing through the 
center of the outflow (left panels) and through the hot core (right panels). 
No obvious difference is seen in the CO(6--5) transition in the two P-V diagrams, while 
the 987~GHz transition shows broader profiles (extending approximately up to -15~km~s$^{-1}$ )
along the axis of the outflow than in the north-south cut through the hot core.  
This is also seen in Fig.~\ref{spectra-out}, where we show  
the 987~GHz and CO(6--5) (averaged over the HIFI beam) spectra at the peak of the redshifted emission: 
the water profile has a clear non-Gaussian redshifted wing and is self-absorbed, while CO(6--5) 
is not and has a broad non-Gaussian profile but no redshifted asymmetry. That the line-profile is broader in water than in CO \citep[generally not seen in other sources using the CO(3--2) line, see][]{vandertak2013} could point to a molecular outflow in an earlier evolutionary phase of SMM2 than of the hot core. Recent observations of low-mass YSOs \citep{2012A&A...542A...8K,2017arXiv170104647M} found that  molecular outflows from class~0 YSOs have more prominent wings in water than those of class~I sources.

The 1113~GHz spectra show additional
absorption features that are due to foreground clouds \citep{vandertak2013}. From single-pointing deep
integration observations of the 1113~GHz line toward the hot core (see Sect. \ref{sec:HIFIanalysis}), at least four features are
detected at about $-16.6, -12.8, -11.4$, and $-3.6$~km~s$^{-1}$.  When
averaging on several pixels, the $-16.6$ ~km~s$^{-1}$ absorption is detected toward SMM8 (and the other positions). The $-12.8$ and $-11.4$~km~s$^{-1}$ absorptions are detected  at SMM1, SMM2, SMM4, SMM7, and SMM8, while the $-3.6$~km~s$^{-1}$ component is not at SMM7 and SMM8. Estimating the exact size of the foreground clouds is impossible with our data: the line-of-sight clouds are mostly seen in absorption only toward the hot core and the other main dust condensations and therefore only toward the continuum emission. We can nevertheless indicate a lower limit of their extent: 20$\arcsec$, 35$\arcsec$, 55$\arcsec$, and 55$\arcsec$  for the foreground clouds at $-3.6, -16.6, -12.8$, and $-11.4$~km~s$^{-1}$, respectively.

In addition to these HIFI maps, the $2_{12}-1_{01}$ line at 179.5 $\mu$m is detected with PACS in absorption over the entire  extent of the IRDC. However, the line is spectroscopically unresolved and no further kinematical information can be derived from the PACS data, whereas the line is spectrally resolved by the HIFI pointed observation toward the hot core. 
Finally, the $3_{03}-2_{12}$ transition at 174.6 $\mu$m, thus involving excited states,  is detected in absorption toward the hot core, then revealing high gas density (see Sect. \ref{kinetic_PACS}). Baseline instabilities prevent us from detecting the line at other positions.

\subsubsection{ H{\sc ii} region}
\label{HIIregion}

\begin{figure}
\centering
\resizebox{\hsize}{!}{\includegraphics[angle=-90]{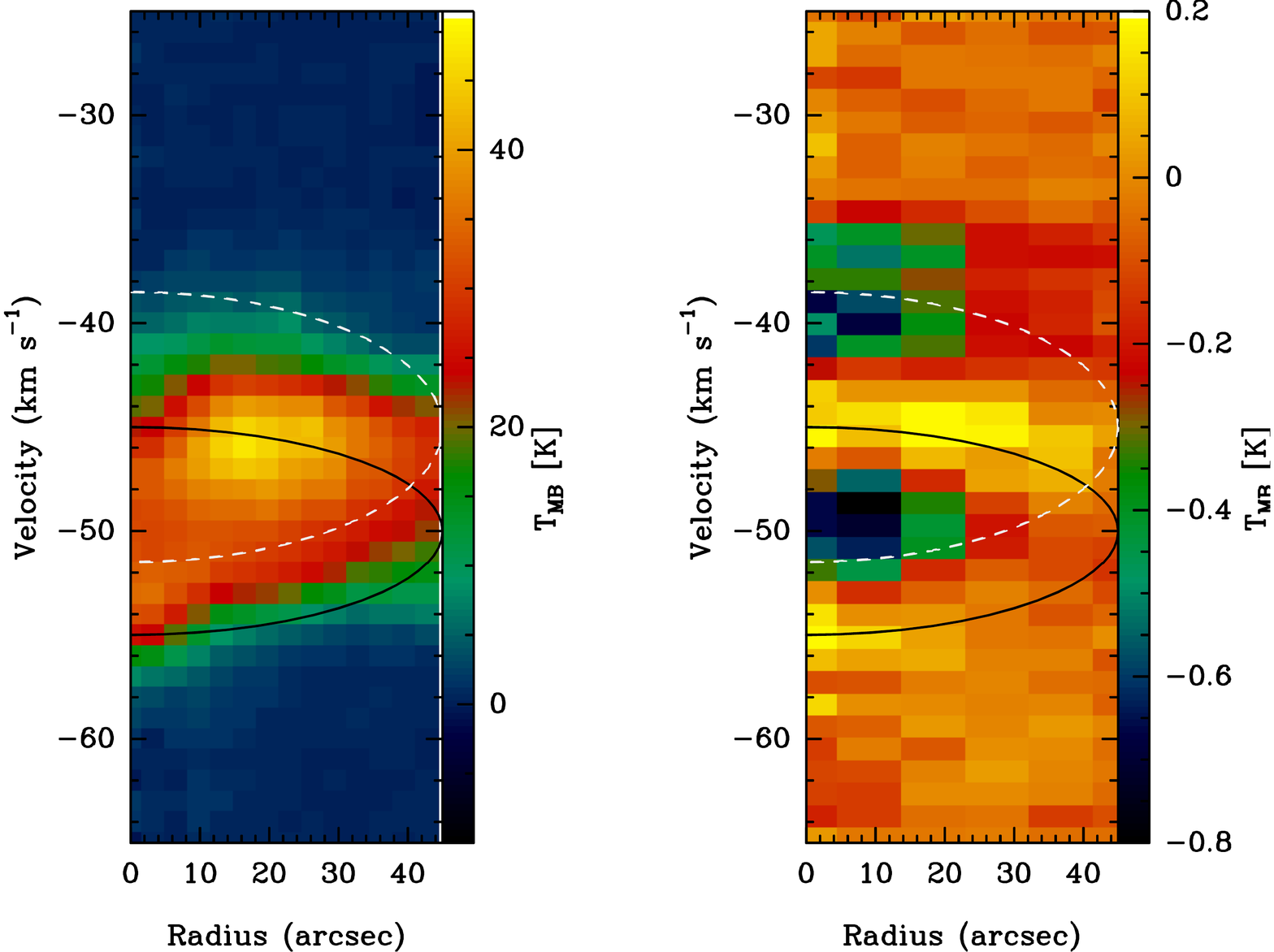}}
\caption{{\it (r-\rm{v})} diagrams of the H{\sc ii} region G327.3--0.5 obtained from the $^{12}$CO(6--5) (left) and from the $1_{11}-0_{00}$-$p$ H$_2$O (right) data cubes. The radius axis is the distance 
to the shell expansion center, chosen to be the peak of the cm continuum emission.  
The black solid half-ellipse represents an ideal shell in  {\it (r-\rm{v})} diagram with an expansion velocity of 5~km~s$^{-1}$ centered on -50~km~s$^{-1}$, the dashed white half-ellipse an ideal shell with an expansion velocity of 6.5~km~s$^{-1}$ centered on -45~km~s$^{-1}$.}\label{rv}
\end{figure}

\begin{figure*}
\centering
\subfigure{\includegraphics[width=9.7cm]{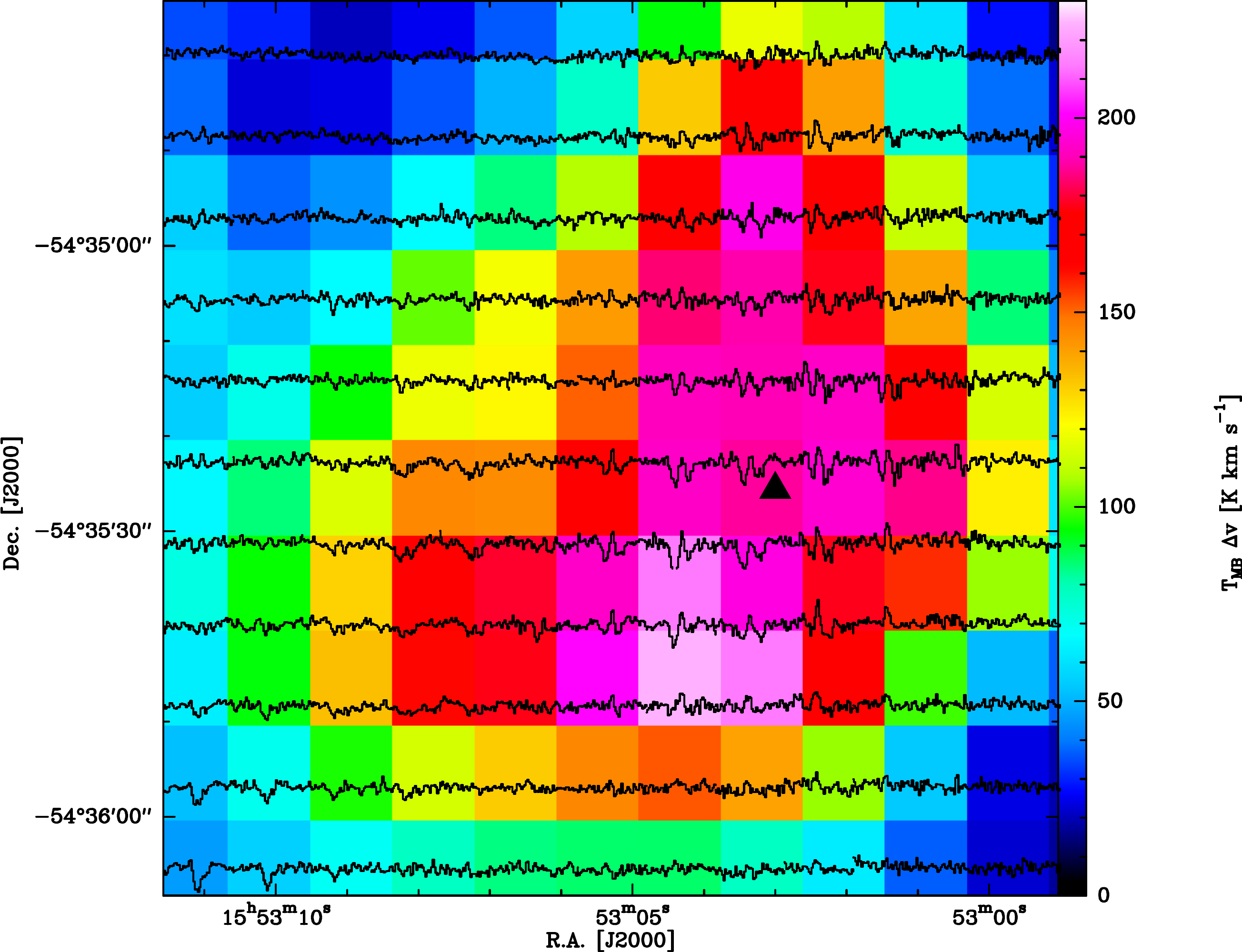}\label{emhii}}
\subfigure{\includegraphics[width=8.1cm]{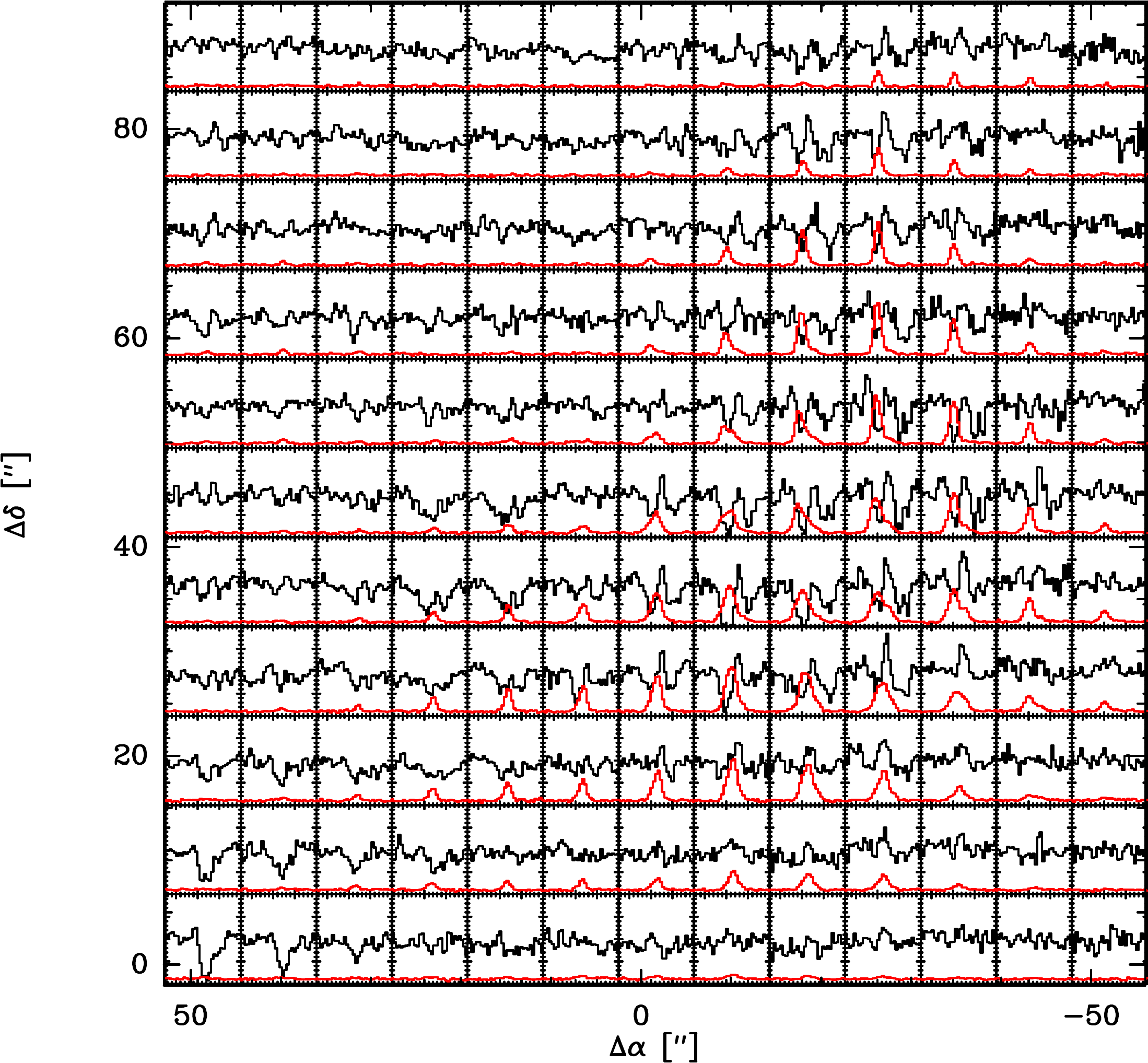}\label{13co109}}
\caption{{\bf Left:} HIFI map of the
$1_{11}\to0_{00}$ p-H$_2$O line toward the H{\sc ii} region overlaid on the $^{13}$CO(6--5) 
integrated emission (see Paper~I).  The temperature axis ranges from -1 to 1.5~K. 
The spectra are continuum subtracted.
The $^{13}$CO(6--5)  data are smoothed to the resolution of the H$_2$O map. {\bf Right:} Spectral maps of the
$1_{11}\to0_{00}$ p-H$_2$O line (black) and of the $^{13}$CO(10--9) transition (red). In both panels, the velocity axis ranges from -65 to -35~km~s$^{-1}$.}\label{hii}
\end{figure*}

The distribution of the 1113~GHz transition in the 
G327.3--0.5 H{\sc ii} region is shown in Fig.~\ref{emhii}, where its spectral map is
overlaid on the integrated intensity of the $^{13}$CO(6--5) line from
Paper~I. The line profile is complex and shows a combination of emission and absorption. Two features
are detected in absorption at $\sim-50$ and
$\sim-38$~km~s$^{-1}$. Interestingly, the emission detected in H$_2$O
is always redshifted compared to the $^{13}$CO(10--9) line (Fig.~\ref{13co109}), which was
observed simultaneously to the 1113~GHz line \citep[presented in][]{2013A&A...550A..10L}. The $^{13}$CO(10--9) seems to be associated with
the absorption at $\sim-50$~km~s$^{-1}$ and peaked at the same velocity as the CO lines observed in \citet{2013A&A...550A..10L}. In Fig.~\ref{rv} we compare the  {\it (r-\rm{v})} diagrams 
of water and $^{12}$CO(6--5). These diagrams  suggest that 
the emission feature at 1113~GHz traces the peak of the $^{12}$CO(6--5) emission.
In Paper~I we speculated that the CO  emission is associated with an expanding shell. 
The two absorption features detected toward the center of the H{\sc ii}
region could be interpreted as due to the back and the front of the
expanding shell. Their separation in km~s$^{-1}$ would be equal
to twice the expansion velocity of the shell. The
  emission feature would be in the direction of the bright borders and would represent 
the mean velocity of the H{\sc ii} region. However, the absolute velocities of water
do not seem to fit those of CO  (see
Fig.~\ref{rv}):  the velocity of the H{\sc ii} region would be around $-45$~km$^{-1}$ and not around $-50$~km$^{-1}$ , as originally suggested 
from the analysis of the CO  isotopologs, and the expanding velocity would be slightly higher (6.5 instead of 5~km~s$^{-1}$).

In the PACS data (see Fig.\ref{fig:PACS_HII_179}), the 179.5 $\mu$m line is detected in absorption
toward all positions where continuum emission is seen, while the
174.6 $\mu$m line is not detected. Additionally, the CH$^+$ (2--1)
transition at 1669.281~GHz is also clearly detected in emission at several positions around the H{\sc ii} region where CH$^+$ traces a photon-dominated region.

\subsection{Pointed observations of the hot core}
\label{sec:HIFIanalysis}

The pointed observations were not performed toward the exact hot core position of G327 (see Sect. \ref{hifi_obs}), but we nevertheless refer to this position as  hot core hereafter. The observed position is 7.5$\arcsec$ west of SMM2 and 16$\arcsec$ northeast of SMM1. As a consequence, the \hoL~and $2_{12}-1_{01}$ (and the \hoM) line observations are missing most of the water around SMM1, while for the other lines both SMM2 and SMM1 are covered by the beam.  

\begin{figure*}
   \begin{minipage}[c]{0.46\linewidth}
     \includegraphics[width=8.cm]{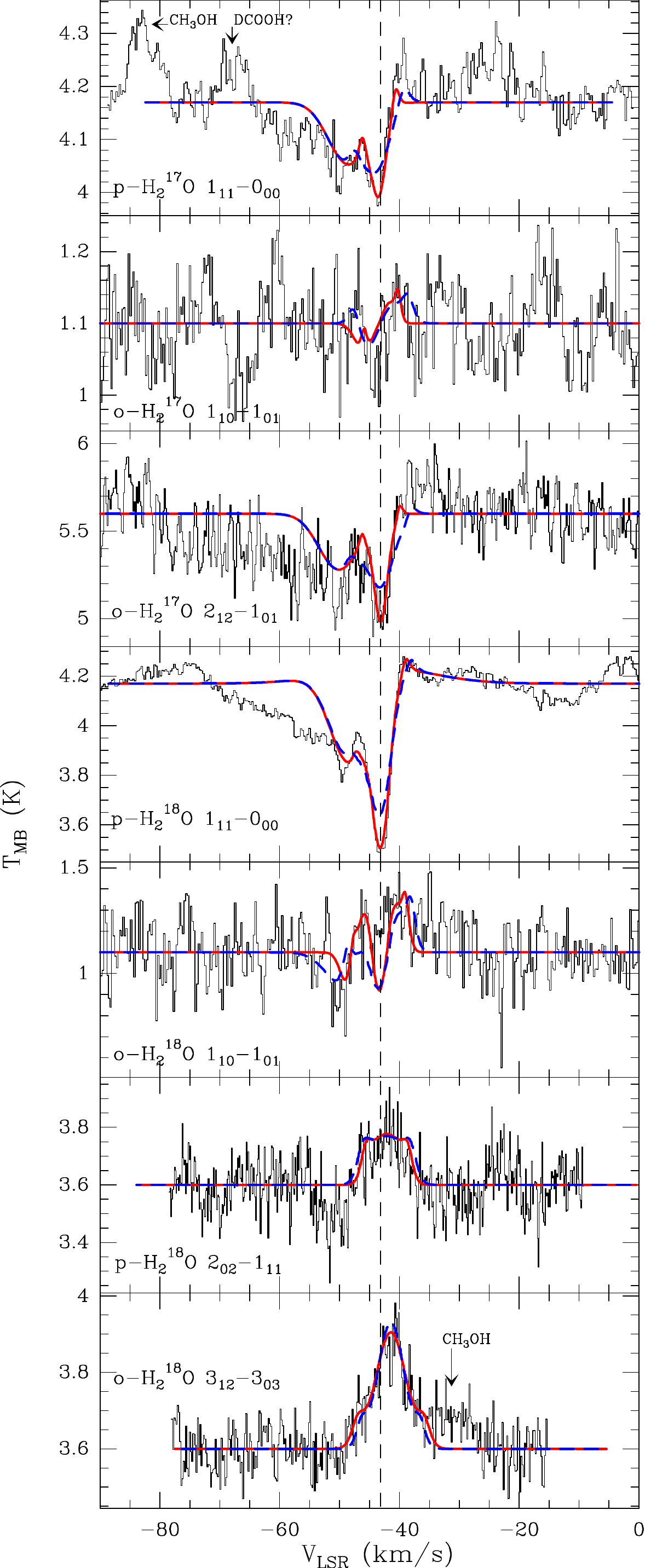}
   \end{minipage} \hfill
   \begin{minipage}[c]{1.96\linewidth}
      \includegraphics[width=8.cm]{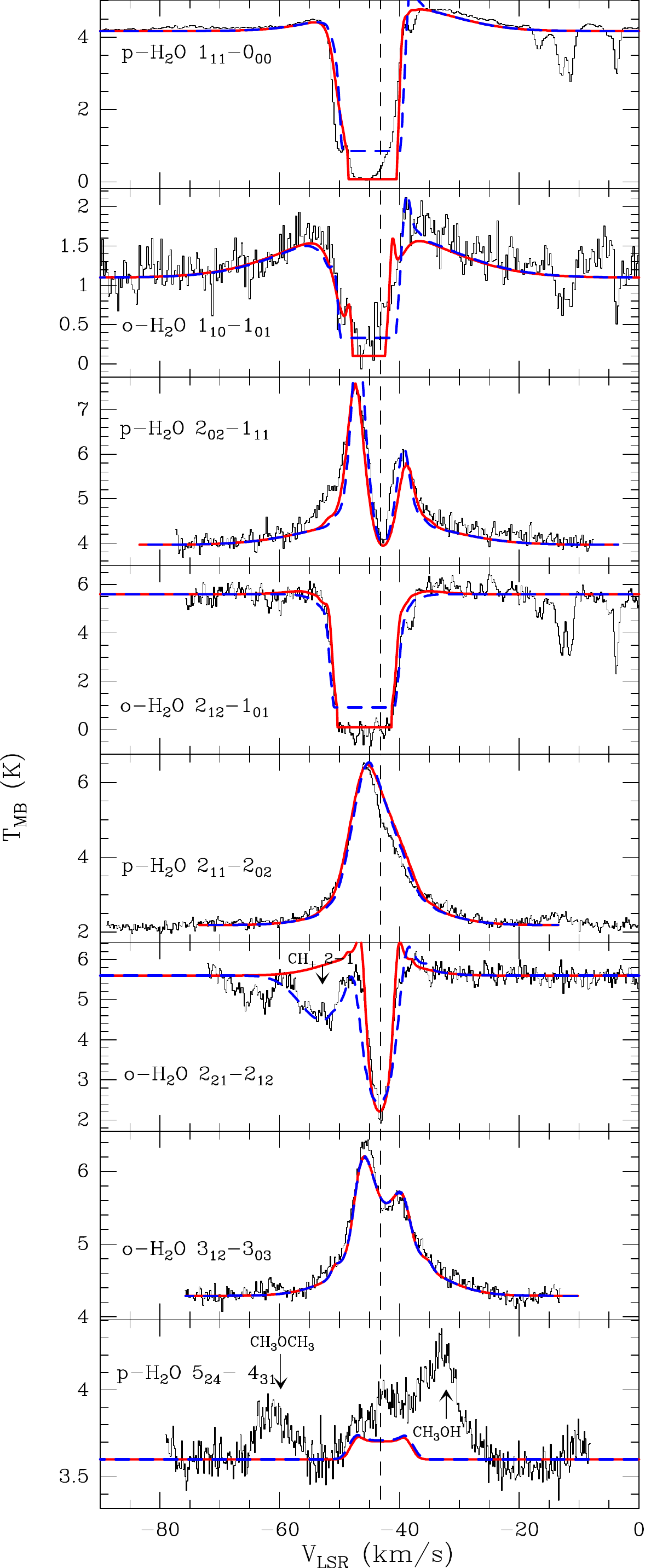}
   \end{minipage}
\caption{HIFI spectra of \watersept$/$\waterhuit~({\em left}) and \water~({\em right}) lines (in black), with continuum for G327.3--0.6 hot core pointed position. The best-fit model with varying (from line to line) and constant (2.6 \kms) turbulent velocity is shown as a red and dashed blue line above the spectra. Vertical dotted lines indicate the \vlsr~(-43.2 \kms~from the line modeling). The spectra have been smoothed to 0.2 \kms, and the continuum is divided by a factor of two.}
\label{FigG327}
\end{figure*}

The spectra including continuum emission are shown in Fig.\ref{FigG327} for the rare isotopologs (\watersept, H$_2^{18}$O) and \water. Spectra of the H$_2$O 1$_{11}$-0$_{00}$ (and H$_2^{18}$O), 2$_{02}$-1$_{11}$, and 2$_{12}$-1$_{01}$ lines have previously been presented by \citet{vandertak2013}. We show the HRS spectra, but for several lines (most of the ground-state lines) we used WBS spectra because the velocity range covered by the HRS was insufficient. For each transition, we derived the peak (emission or absorption dip) main-beam and continuum temperatures, half-power line widths for the different line components from multi-component Gaussian fits, made with the CLASS software, and opacities for lines in absorption (line parameters are given in Table~\ref{table_param}).  

Several foreground clouds \citep[][]{vandertak2013} contribute to the spectra in terms of water absorption at V$_{lsr}$ (-3.7, -11.4, -13, -16.6 \kms) shifted with respect to the source velocity in the \hoC, \hoK, \hoN~and \hoI~line spectra.

The velocity components are attributed to cavity shocks and envelope component as for low-mass (LM) protostars \citep[][]{mottram2014} or for other high-mass studies \citep[see][]{herpin2016}. The broad (FWHM$\simeq$20-35 \kms) velocity component arises in cavity
shocks\textup{} (i.e., shocks along the cavity walls) as its narrower version, the medium component (FWHM$\simeq$5-10 \kms), coming from a thin layer (1- 30 AU) along the outflow cavity where non-dissociative shocks occur. The envelope component (narrow component with FWHM $<$5 \kms) is characterised by small FWHM and offset, that is, emission from the quiescent envelope.

In the following we refer to the commonly assumed hot core velocity of $\sim -45$\kms \citep[][]{bisschop2013}, but (APEX) observations of rare CO isotopolos instead point to lower velocities: -43.7, -44.3, and -44.7 \kms~for C$^{18}$O J=8-7 (and $^{13}$CO 10-9), 6--5, and $^{13}$CO 6--5, respectively \citep[]{2011A&A...527A..68R,2013A&A...550A..10L}. Interestingly, the higher excitation lines tend to be more blueshifted. 

\begin{table*}
  \caption{Observed line emission parameters for the detected lines with HIFI toward G327-0.6 hot core pointed position. $\varv$ is the Gaussian component peak velocity. $\Delta\varv$ is the velocity full width at half-maximum (FWHM) of the narrow, medium, and broad components. The opacity $\tau$ is from absorption lines.}
\begin{center}
\label{table_param}     
\begin{tabular}{lccccccccc} \hline \hline
 Line  & $T_{mb}$ & $T_{cont}$ & $\varv_{nar}$ & $\Delta\varv_{nar}$ & $\varv_{med}$ & $\Delta\varv_{med}$ & $\varv_{br}$ & $\Delta\varv_{br}$  & $\tau$ \\ 
 & [K] & [K] & [\kms]  & [\kms] & [\kms] & [\kms] &  [\kms] &[\kms]  & \\ 
\hline                        
   \hoA     &  0.88 & 1.1  & -49.8$\pm$0.3$^a$ & 2.4$\pm$0.4&  & & & & 0.22$\pm$0.05\\ 
   \hoF     &  3.95 & 3.6 & $-51.0\pm0.3^a$ & $4.4\pm0.4$& -42.4$\pm$0.3 & 6.4$\pm$0.7& & & \\ 
   \hoG   & 4.67 & 4.29  & & & -41.7$\pm$0.3& 5.4$\pm$0.7& &  &  \\
   \hoI$^b$     & 3.50 & 4.17  & -43.2$/$-49.4$\pm$0.2$^a$& 3.0$/$2.2$\pm$0.3$/$0.5 &  & & -54$\pm1^a$ & 20$\pm1$& 0.17$\pm$0.06\\
   \hoJ$^b$    & 3.95 & 4.17  & -44.1$\pm0.2^a$& 3.2$\pm0.4$&  -49.1$\pm0.6$ & 5$\pm$1& -53$\pm1^a$ & 20$\pm2$ & 0.05$\pm$0.02\\
   \hoM   & 4.95 & 5.6  & -43.2$\pm$0.2$^a$ & 3.1$\pm$0.4 & -40.1$\pm0.9^a$& 10$\pm2$ & -54$\pm3^a$& 21$\pm2$ & 0.12$\pm$0.05 \\
\hline
   \hoC   & 0.04 & 1.1  & &  & -45.7$\pm$0.2$^a$& 8.3$\pm$0.4& -43.7$\pm$0.7 & 30$\pm$2 & 3.3$\pm$0.7\\
   \hoD      & 6.52 & 2.19  & -43.3$\pm0.1^a$& 3.1$\pm0.1$ &-44.8$\pm$0.1 & 6.4$\pm$0.2 & -42.8$\pm$0.2& 18.4$\pm$0.6&\\
   \hoP      & 4.0 & 3.6  & & & -42.1$\pm$0.2 & 10.$\pm$0.5 & &   & \\
   \hoE      & 7.71 & 3.96  & -43.3$\pm$0.1$^a$ & 4.0$\pm$0.1& -44.5$\pm$0.1 & 8.8$\pm$0.2 & -42.2$\pm$0.4 & 33$\pm$1 & \\
  \hoH   & 6.45 & 4.29  & -43.0$\pm0.1^a$ & 3.4$\pm0.1$ & -43.9$\pm$0.1& 5.6$\pm$0.2 & -42.1$\pm$0.2 & 15.9$\pm$0.3 &\\
   \hoK$^b$    & 0.06 & 4.17  & -43.2$\pm0.2^a$ & 4.9$\pm0.2$ & -48.2$\pm$0.2$^a$ & 5.4$\pm$0.2 & -41.5$\pm$2 & 23.7$\pm$0.6 & 4.$\pm$1\\
   \hoL$^b$     & 2.1 & 5.6  & -43.1$\pm$0.2$^a$& 3.4$\pm$0.1& &  & && 1.0$\pm$0.4 \\
   \hoN$^b$    & 0.01 & 5.6 & -42.5$\pm0.2^a$ & 4.8$\pm0.2$ & -48.1$\pm$0.2$^a$& 5.8$\pm$0.2& & & 6$\pm$1\\
\hline
\end{tabular}
\end{center}
\tablefoot{$^a$ in absorption, $^b$ WBS data}
\end{table*}

\subsubsection{Rare isotopologs}

The para ground-state line of \watersept~and \waterhuit~(see Fig. \ref{FigG327}) is detected and exhibits the same line profile in absorption, made of an envelope component (FWHM$\sim$3 \kms) slightly blueshifted (less than 1 \kms) from the APEX $V_{LSR}$, one narrow$ \text{ or }$medium redshifted component in absorption, and a broader absorption that is more redshifted (by $\leq$10 \kms). This broad absorption is discussed in Sect. \ref{sec_dyn}. A similar profile is observed for the \hoM~line. While the \hoB~line is not detected, the \hoA~line is tentatively detected with a weak and narrow absorption at -49.8 \kms. 

In contrast, a relatively strong signal is observed for the \hoF~and \hoG~transitions (see Fig. \ref{FigG327}) showing the same cavity shock blueshifted component in emission. In addition, the \hoF~line exhibits an absorption at -51 \kms similar to the one observed for the ground-state lines. We note that the \hoG~line is blended with a CH$_3$OH line.

The absorption that is either narrow (\hoA, \hoF, \hoI), medium (\hoJ), or broad (\hoI, \hoJ, \hoM) observed at velocities between -49.1 and -54 \kms~is most likely the broad absorption component seen in the NH$_3$ line profile (at -49.62 \kms~with $\Delta v \simeq 11.1$ \kms) by \citet{wyrowski2016} and could be due to absorption by foreground material (see Sect \ref{sec_abun} for a detailed discussion). 

\subsubsection{Water lines}

All targeted \water~lines have been detected in absorption for the ground-state and the 2$_{21}$-2$_{12}$ lines, while other transitions exhibit a line profile in emission with some self-absorption at the source velocity. One line, \hoP, is in pure emission (cavity shock component), but is blended with a methanol line.   

An envelope component in absorption is seen in all lines but the \hoP~and \hoC transitions, centered at $\sim$-43 \kms. The medium cavity shock component is observed in absorption for the ground-state lines, redshifted by 2-4 \kms, while it is seen
in emission for the other water line and roughly at the source velocity. In addition, a broad component (up to 30 \kms) is seen in emission in most of the lines (Sect. \ref{sec_dyn}) and is blueshifted.

All \water~lines in absorption are optically thick based on line$/$continuum ratios (with opacities between 1 and 6, see Table \ref{table_param}). The optically thick \hoE, \hoD, and \hoH~lines are strongly blue asymmetric, that is to say, they exhibit inverse P-Cygni profiles, hence they probably indicate infalling material.  

\subsubsection{Carbon species}

\begin{figure}
\centering
\resizebox{\hsize}{!}{\includegraphics{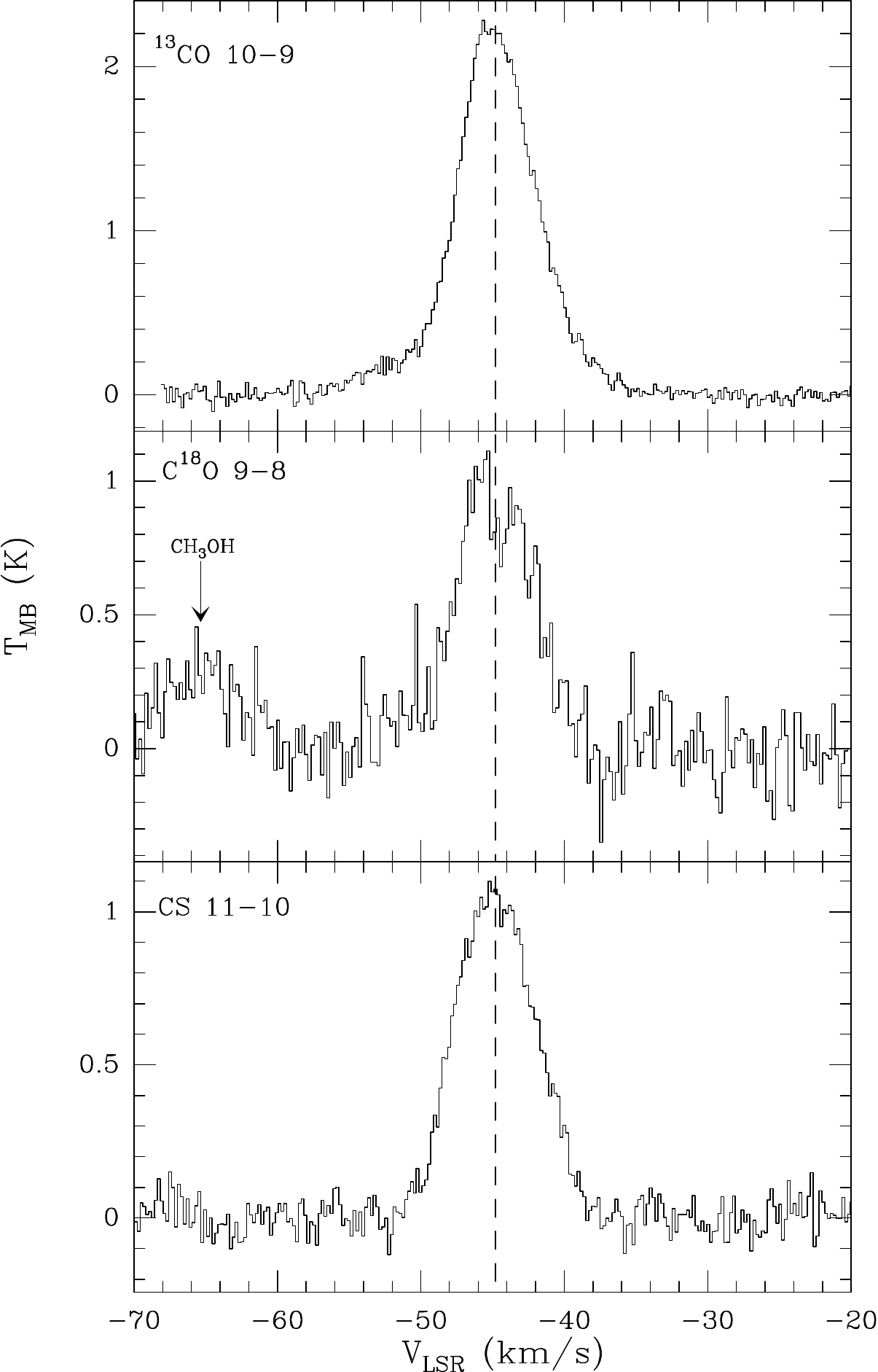}}
\caption{HIFI spectra of $^{13}$CO J=10--9, C$^{18}$O J=9--8, and CS J=11--10 lines for the HIFI pointed position. The spectra have been smoothed to 0.2 \kms. The vertical dotted line indicates the source velocity derived from Gaussian fitting, i.e., --44.8 \kms.}
\label{Fig_C}
\end{figure}

In addition to the water lines, a few lines from carbon species have been observed and are shown in Fig.\ref{Fig_C}: $^{13}$CO J=10--9, C$^{18}$O J=9--8, and CS J=11--10. These three lines are in emission and centered at --44.8 \kms, hence at a slightly redshifted velocity compared to what derived \citet{2011A&A...527A..68R} and \citet{2013A&A...550A..10L} from ground observations. Line profiles exhibit a cavity shock component of 5.3-6.5 \kms, but a broader component (FWHM$\sim$11 \kms) is also observed for the  $^{13}$CO J=10--9 line. 

\section{Analysis}

\subsection{Water abundance from the HIFI data}\label{hifi}
The opacity of a spectrally resolved unsaturated absorption line can be determined by\begin{equation}\label{tau}
\tau = -{\rm ln}\biggl[\frac{T_{\rm L}}{T_{\rm C}}\biggr]
,\end{equation}
where $T_{\rm L}/T_{\rm C}$ is the line-to-continuum ratio. In this case, the column density of the absorbing species can be derived by (for ground-state lines, assuming negligible excitation)

\begin{equation}\label{colden} 
N_{\rm tot} = \frac{8 \pi \nu^3}{A_{\rm ul} c^3} \Delta v \frac{g_{\rm l}}{g_{\rm u}} \tau
.\end{equation}

In the case of the 1113~GHz transition, the absorption is saturated
toward all positions in the IRDC. In addition, the corresponding
H$_2^{18}$O line (observed in the same setup as the main isotopolog
line) is not detected in the HIFI maps. Therefore, the opacity of the
1113~GHz line cannot be computed analytically from Eq.~\ref{tau}.  In
this case, the optical depth can be derived from a curve-of-growth
analysis, once the equivalent width, $W$, of the transition is
computed. We have
\begin{equation}
\label{ }
W = \int{\kappa(\nu) d\nu}
,\end{equation}

where $\kappa(\nu$ is the absorption coefficient. We solved it graphically with a normal curve-of-growth analysis (log(W) vs log(tau)).

\begin{figure}
\centering
\resizebox{\hsize}{!}{\includegraphics{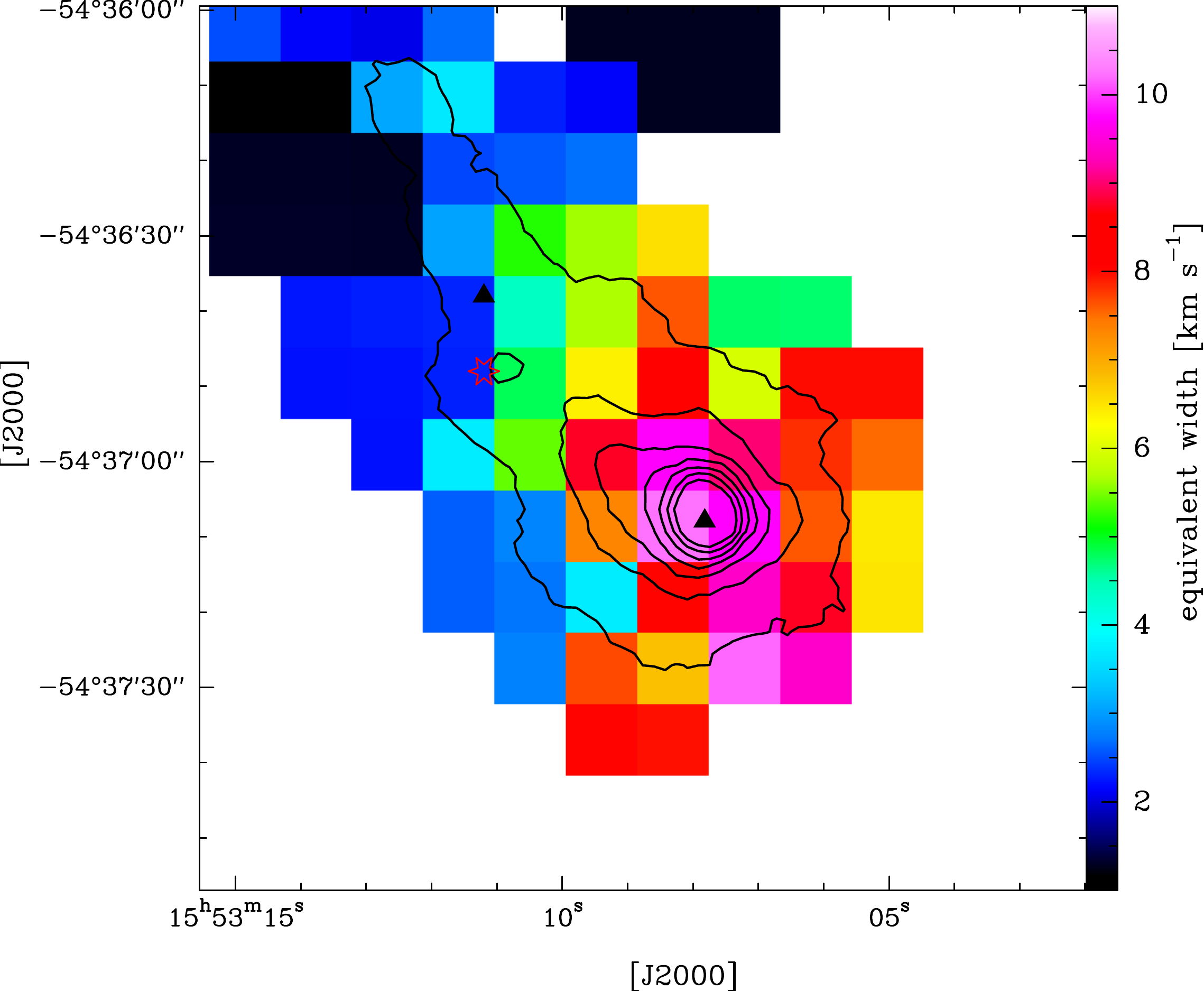}}
\caption{Distribution of the equivalent width of the $1_{11}\to0_{00}$ $p-$H$_2$O line in the IRDC. The black contours show the
  SABOCA continuum emission at 350 $\mu$m from 5\% of the peak flux in steps of 10\%. }\label{ew}
\end{figure}

The distribution of $W$ in the IRDC is shown in Fig.~\ref{ew}. Results toward the H{\sc ii} region
are not reliable given the complex line profile of the 1113~GHz transition in this part of
the source (see Fig.~\ref{emhii}). In the IRDC,  $W$
decreases steeply from a value of $\sim 10$~km~s$^{-1}$ toward the
hot core down to a value of 5~km~s$^{-1}$ at the edge of the
IRDC. 

To derive the opacity of a transition from a curve-of-growth analysis,
the line profile must be known. Line profile and line width of the
1113~GHz transition cannot be inferred from our data as the line is
highly saturated.  As first approximation, we can assume that the
1113~GHz line has the same profile and line width as the
C$^{18}$O(3--2) transition, which has a low opacity and a low energy
($E_u\sim32$~K) and therefore is likely to trace the same cold
component that absorbs water.  We note that the width of H$_2$$^{18}$O cannot be used because the rare isotopolog line is not detected out of the central region. C$^{18}$O(3--2) was observed by
\citet{2006A&A...454L..91W}, and it has Gaussian profiles and typical
widths of 4~kms$^{-1}$ at the IRDC position and of 6~kms$^{-1}$ at the
hot core. For comparison, the width of the narrow component of the H$_2$$^{18}$O $1_{11}-0_{00}$ line is around 3 \kms at the hot core position, but the line profile also exhibits a broad component in absorption in the blue part that is due to the outflow. At the hot core position, the equivalent width of the
1113~GHz line is $\sim 10$. This value corresponds to an optical depth
of 80 for $\Delta {\rm v}=6$~kms$^{-1}$. However, for this value of
$W$, the results are strongly dependent on the adopted line width and
vary between 70 and 130 for $\Delta {\rm v}$ in the range
5--7~kms$^{-1}$ , with higher opacities corresponding to narrower line
widths.  At the IRDC position, $W\sim 5$ corresponds to an optical
depth of $\sim15$, and it does not substantially change with the line
width. The validity of these estimates can be cross-checked on the
hot core, where the deeper single-point observations of the 1113~GHz line 
allow us to detect the H$_2^{18}$O and H$_2^{17}$O isotopologs of
the line. From Eq. 1, we derive an optical depth of $\sim$0.16 for the $1_{11}-0_{00}$ p-H$_2^{18}$O line, and of $\sim$0.05 for the $1_{11}-0_{00}$ p-H$_2^{17}$O line. These values translate into  lower limits of 62-87 for the optical depth of the main isotopolog line, in agreement
with the result from the curve-of-growth method, assuming $^{16}$O/$^{18}$O=390 and $^{18}$O/$^{17}$O=4.5, respectively \citep{wilson1994}.

Equation~\ref{colden} translates (assuming an o$/$p ratio of 3) into a total column density of water
of $1\times10^{15}$~cm$^{-2}$ for the hot core for an optical depth of 80 and
a line width of 6\,km\,s$^{-1}$. At the IRDC position, the column density of H$_2$O is $2\times10^{14}$\,cm$^{-2}$ for $\tau=15$ and $\Delta\varv=5$\,km\,s$^{-1}$.  In Paper~I, we derived the distribution of the
H$_2$ column density over the whole region from CO and $^{13}$CO(6--5)
maps. We can assume that CO and $^{13}$CO(6--5) trace the same gas that
is absorbing the 1113~GHz transition and derive the abundance of water
in the region.  Toward the hot core, the H$_2$ column density is
3.0$\times 10^{22}$\,cm$^{-2}$, while it decreases to $1.0\times 10^{22}$\,cm$^{-2}$
at the IRDC position\footnote{In Paper~I we used an averaged value of 60 for the $^{12}$CO/$^{13}$CO abundance, while here we adopt a value of 53 for a galactocentric distance of 6\,kpc}. These values correspond to abundances of water
relative to H$_2$ of 3$\times 10^{-8}$ toward the hot core, and
$2\times 10^{-8}$ at the IRDC position. However, the uncertainties on
these values are large, especially at the hot core position, where the
strong saturation in the 1113\,GHz line does not allow a precise
determination of $N_{\rm {H_2O}}$.  Given the inferred range of
opacities of the 1113\,GHz line at this position, the abundance of
water could vary between 3$\times 10^{-8}$ (for $\Delta\varv=4$\,km\,s$^{-1}$) and 8$\times 10^{-8}$ (for $\Delta_\varv=7$\,km\,s$^{-1}$).

Our observations suggest that the abundance of water does not  change along the 
IRDC with values close to a few times 10$^{-8}$. This abundance is consistent with
results from several studies toward the outer part of envelopes around massive YSOs 
and in  the foreground clouds \citep[e.g., ][]{2000ApJ...539L..97S,2002ApJ...581L.105B,2013ApJ...765...61E}, and with chemical models \citep[e.g.,][]{2002A&A...389..446D}. 

\subsection{Kinetic temperature from PACS data}\label{pacs_res}
\label{kinetic_PACS}
The PACS data suffer from two problems that make their analysis
difficult: low spectral resolution and flux leakage. The latter arises because the PACS spaxels have a projected size of
9\farcs4$\times$9\farcs4 on the sky, while the point-spread function of
the {\it Herschel} telescope at $179.5\mu$m is approximately
12\farcs6. For a point-like source that is perfectly centered on one spaxel,  44\% of the flux is recovered
at 179 $\mu$m (Fig.~7 of the PACS spectroscopy performance and calibration document\footnote{http://herschel.esac.esa.int/twiki/pub/Public/PacsCalibrationWeb/\\PacsSpectroscopyPerformanceAndCalibration\_v2\_4.pdf}).
However, the flux loss depends on the source structure for extended
sources. In this case, the fraction of flux seen by PACS can be
inferred only by deconvolving a source model image by the PACS point-spread function and comparing the flux seen in this {\it \textup{synthetic}}
observations with the original one in the model.  This procedure
should be performed for the continuum emission and for the
absorption/emission of each transition separately, as the fraction of
the flux recovered by PACS depends on the structure of the source in
that particular tracer. Unfortunately, for G327.3--0.6 we do not have
any continuum image at higher angular resolution than that of the PACS
data to input as source model, nor we have any detailed knowledge of
the distribution of water. Moreover, the approach described by
\citet{2012A&A...540A..84H} and \citet{2013A&A...552A.141K} of summing
up fluxes from adjacent spaxels is impractical in our case as
different spaxels most likely see different sources given the complexity
and the distance of the region. 

Since the 179.5 $\mu$m and the 174.6 $\mu$m water lines are seen in absorption, the  
line-to-continuum ratio would not be affected by flux leakage if the two transitions 
had the same distribution of the continuum emission at the corresponding wavelength. We could 
test this
assumption at 179.5 $\mu$m toward the  position observed with HIFI in the $o$-H$_2$O $2_{12}-1_{01}$ line. This coincides with the position reported
by \citet{1992PhDT.......252B}, shifted by approximately
(8\farcs7,5\farcs5) from the current hot core position. The comparison
between the PACS and the HIFI 179.5 $\mu$m line at this position infer a
difference of about 10\% between the equivalent width of the water
line obtained in the same velocity range, which value is indeed
comparable to the relative calibration error between the two
instruments. This test confirms that the PACS
continuum and spectral observations at 179.5 $\mu$m are affected by
flux loss in a similar way, but it is impossible to quantify
this.

Assuming that there is no flux leakage, from the absorption of $o$-H$_2$O $3_{03}-2_{12}$ line at 174.6 $\mu$m,
we can estimate an upper limit to the excitation temperature of the
line, as this must be lower than the continuum temperature at 174.6 $\mu$m. The continuum level
toward the hot core is approximately 1300~Jy/spaxel, which
corresponds to a brightness temperature of 3.6~K, using a beam size of
12\farcs3. At these wavelengths, and for typical temperatures of star-forming regions, the Rayleigh-Jeans approximation is not valid anymore,
and the
(exact) Planck brightness temperature is 21~K, implying a lower 
excitation temperature for the $o$-H$_2$O $3_{03}-2_{12}$ line. Since
the critical density of this transition is very high, its excitation
temperature depends mostly on the kinetic temperature of the medium
and on the column density of ortho water. 
According to the RADEX online radiative transfer code \citep{2007A&A...468..627V}, 
for a  column density
of $10^{14}$~cm$^{-2}$ and a line-width of 6~km~s$^{-1}$ (see Sect.~\ref{hifi}), 
an upper limit of 20~K for the excitation
temperature of $3_{03}-2_{12}$ line implies an upper limit to the
kinetic temperature of 40~K, in agreement with the excitation temperature of 30-35~K for the hot core inferred in Paper~I from CO(6--5) observations. 
The upper limit to the kinetic temperature of the gas
increases with column density, and corresponds to 30~K for $N_{o-\rm{H_2O}}=10^{13}$~cm$^{-2}$  
and to 150~K for $N_{o-\rm{H_2O}}=5\times10^{14}$~cm$^{-2}$.

\section{Modeling of the HIFI lines}
\label{sec:model}

This section intends to model the full line profiles in a single spherically symmetric model with different kinematical components that are due to turbulence, infall and outflow.

\subsection{Method}
\label{method}

The envelope temperature and density structure for the hot core from \citet{vandertak2013} is used as input to the 1D radiative transfer code RATRAN \citep[][]{hogerheijde2000} in order to simultaneously reproduce all the water line profiles, following the method of \citet{herpin2012}. The H$_2$O collisional rate coefficients are from \citet{daniel2011}. We assume a single source within the HIFI beam throughout our analyis, but we know   that this source consists of two objects (SMM1, $\sim3770$\Msol, and SMM2, $\sim$200\Msol, see \citealt{2009A&A...501L...1M} and Sect.\,\ref{cont_emi}), separated by $\sim$ 23\arcsec, and that our observations are pointing between these two objects (see Sect. \ref{hifi_obs}). The insufficient knowledge of the SED for each of these subsources and the lack of spectral information prevent any more detailed modeling of the structure. Since SMM1 is $\sim20$ times more massive than SMM2, we may assume that the emission is dominated by SMM1.

Adopting here a single-source structure that encompasses the substructure within the HIFI beam, the source model has two gas components: an outflow and the protostellar envelope. The outflow parameters, intensity, and width come from the Gaussian fitting presented in Sect. \ref{sec:HIFIanalysis} for the broad component. The envelope contribution is parametrized with three input variables: water abundance ($\chi_{H_2O}$), turbulent velocity ($V_{tur}$), and infall velocity ($V_{inf}$). The width of the line is adjusted by varying $V_{tur}$. The line asymmetry is reproduced by the infall velocity. The line intensity is best fit by adjusting a combination of the abundance, turbulence, and outflow parameters. We adopt the following standard abundance ratios (same ratios for all the lines):  4.5 for \waterhuit$/$\watersept~\citep[][]{thomas2008}, and 3 for ortho$/$para-H$_2$O. Based on \citet{wilson1994}, the \water$/$\waterhuit~abundance ratio is assumed to be 390. The model assumes a jump in the abundance in the inner envelope at 100 K because the ice mantles evaporate. All details about the method are given in \citet{herpin2012}.

\subsection{Abundance and kinematics results}
\label{sec_abun}

The analysis presented in Sect.\ref{sec:HIFIanalysis} has shown that the width of the velocity components is not the same for all lines (e.g., half-power line widths from 2.4 to 4.9 \kms~for the narrow component, see Table \ref{table_param}). As a consequence, a model with equal velocity parameters for all lines does not fit the data well. A turbulent velocity of 1.5 and 2.1 \kms~ for the \watersept~and \waterhuit~ground-state lines in absorption also gives a good result for the \water~lines in absorption. In contrast, a higher turbulent velocity of 2.6-3 \kms~is needed for the lines in emission. We note that from RATRAN modeling of the NH$_3$ $3_{2+}-2_{2-}$ line, \citet{wyrowski2016} derived a turbulent velocity of 2.3 \kms. We then ran a model with a constant turbulence of 2.6 km\,s$^{-1}$ for all lines (the best compromise we obtained after exploring a range of values). We also tested two other options that do not improve the fit significantly (see Fig. \ref{FigG327}): the first option is a turbulence varying with radius following \citet{herpin2012} and \citet{herpin2016}, the second option adjusts the turbulence line by line based on what is observed. The limiting factor rather seems to be the assumed spherical symmetry.

All modeled lines are centered at roughly $-43\pm0.5$ \kms. The infall velocity is estimated to be -3.2 \kms~(at $\sim$1500 AU). A foreground absorption was included at $-48\pm0.5$ \kms~ with a width of 7.5 \kms~ for the ground-state water lines, but this component has no effect on the water abundance, very likely because it is sufficiently far from the source velocity. What we see is the absorption of the blue part of the outflow. This broad absorption has been observed and described in \citet{herpin2016} for high-mass protostellar sources NGC6334IN and DR21(OH). Interestingly, this absorption is at a similar velocity to one of the absorption features observed in the H{\sc ii} region (see Sect. \ref{HIIregion}) and could be the same cloud over the whole region (see discussion in the next section). We did not try to reproduce the outflow absorption seen in \hoI~and \hoJ~lines.

The water abundance is constrained by the modeling of the entire set of observed lines. Only a few lines (\hoG, \hoP~and \hoH) are optically thin enough to probe the inner part of the envelope, part of all water line profiles is produced by water excited in the inner part and is revealed by the high spectral resolution of these observations. The \water~abundances relative to H$_2$ are $5.2\times10^{-5}$ in the inner part where $T>100$ K, while the outer abundance (where $T<100$K) is $4\times10^{-8}$ (we estimate the uncertainty to 30\%), consistent with what we found in Sect. \ref{hifi} for this position. No deviation from the standard $o/p$ ratio of 3 is found.

\section{Discussion}
\label{sec_disc}

\subsection{Source structure and dynamics of the hot core region}
\label{sec_dyn}

\begin{figure}
\centering
\resizebox{\hsize}{!}{\includegraphics{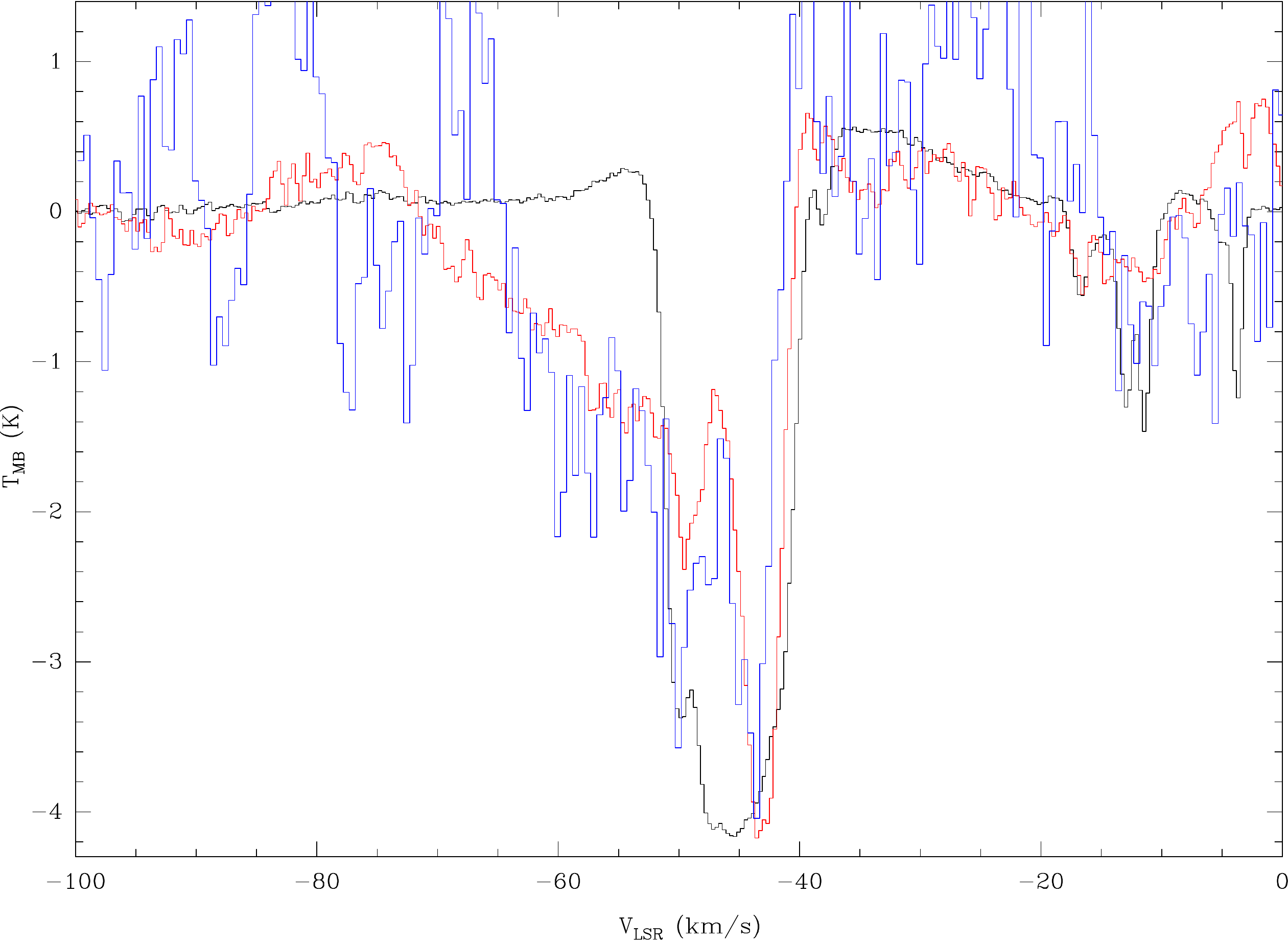}}
\caption{Spectra of \hoI~ ($\times 6.2$, in red) and \hoJ~($\times 21$, in blue) lines overplotted on the \hoK~spectrum (in black). The spectra have been smoothed to 0.3 \kms~(1.4 \kms for the \watersept~spectrum).}
\label{FigCompare}%
\end{figure}

\begin{figure}
\centering
\resizebox{\hsize}{!}{\includegraphics{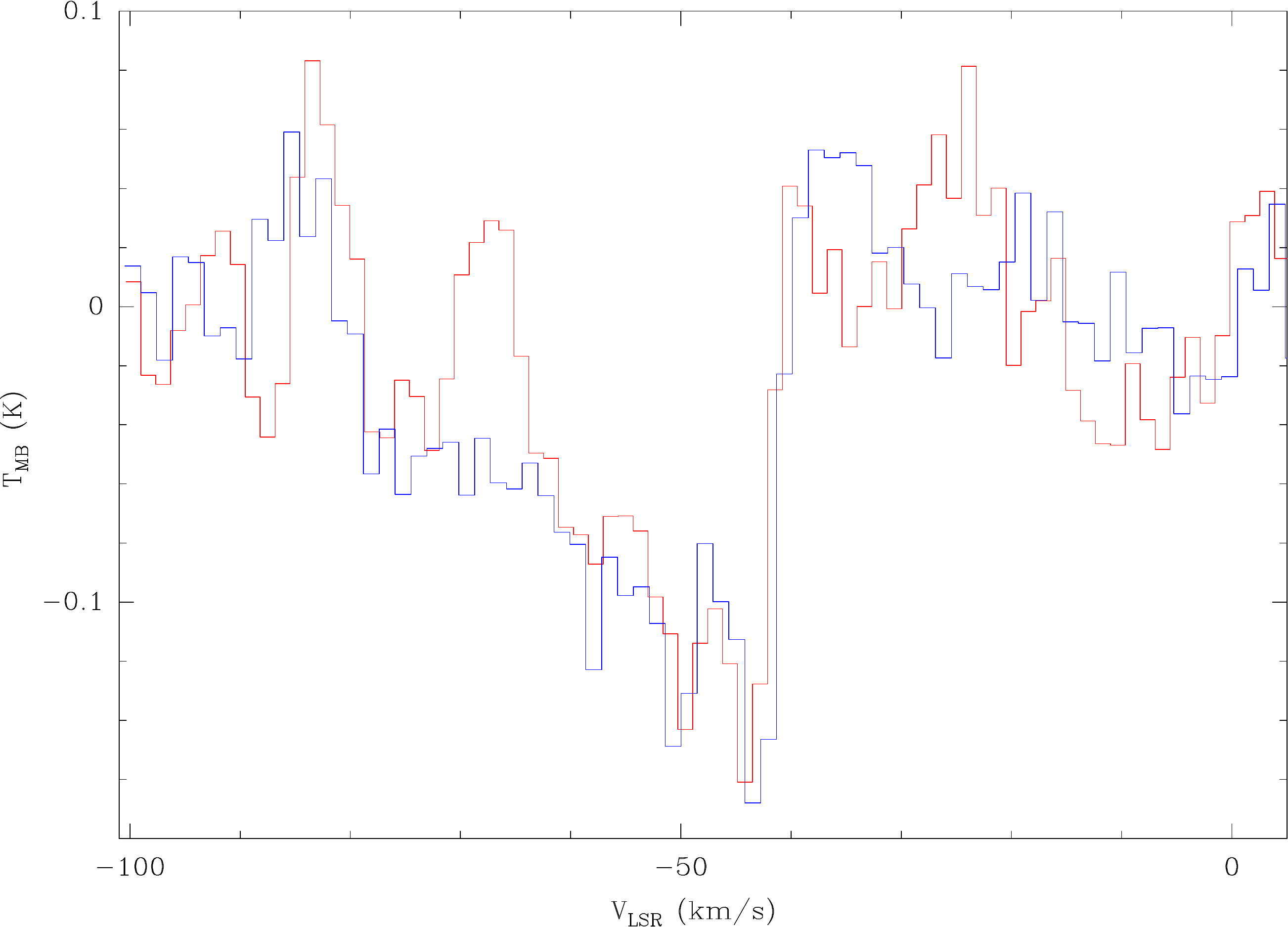}}
\caption{Spectra of \hoM~($\times 0.3$, in blue) line overplotted on the \hoJ~spectrum (in red). The spectra have been smoothed to 1.4 \kms.}
\label{FigCompare2}%
\end{figure}

\begin{figure}
\centering
\resizebox{\hsize}{!}{\includegraphics{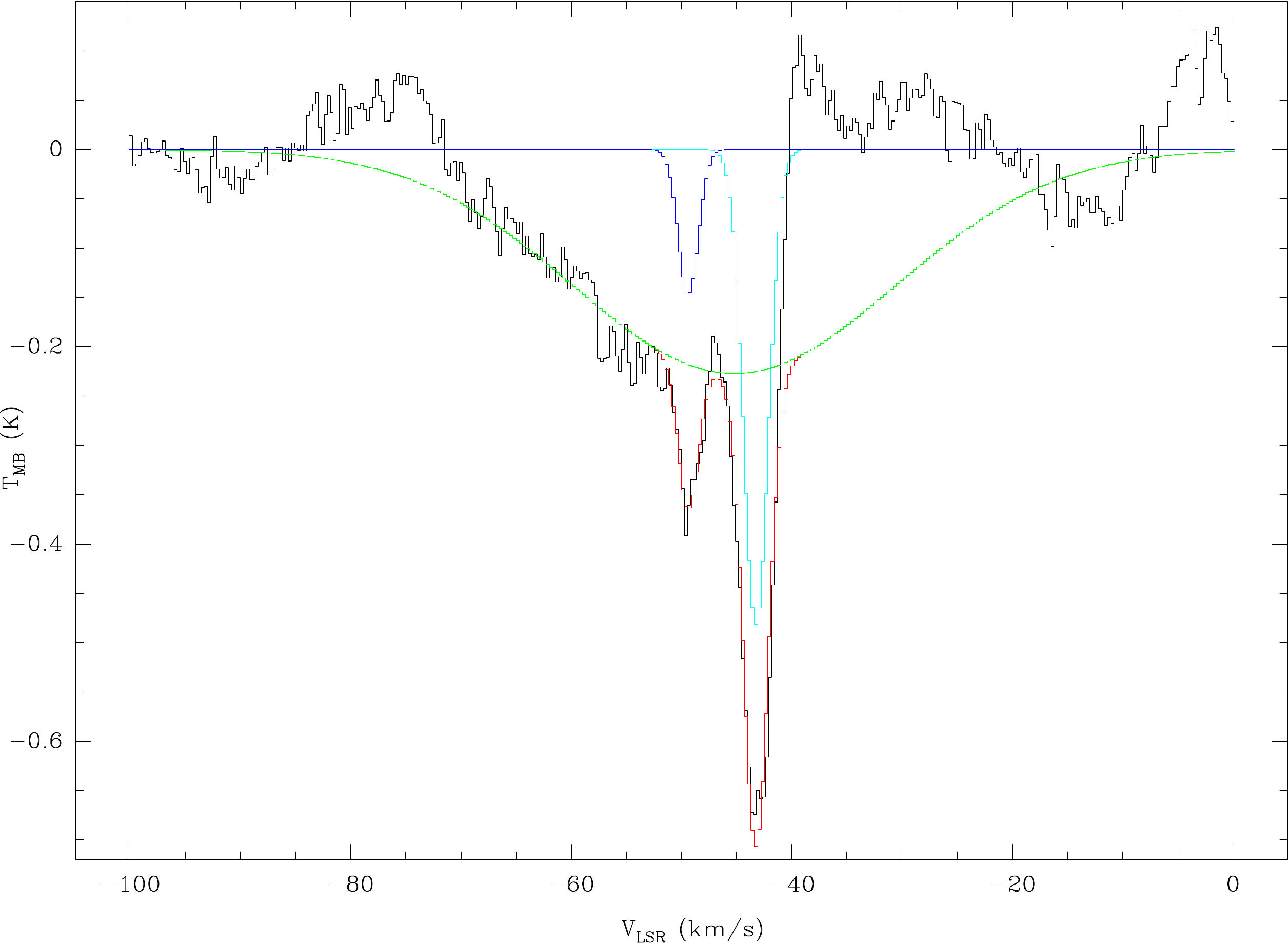}}
\caption{Spectra of \hoI~line (in black) showing the different Gaussian components used to fit the line (red= green+blue+purple).}
\label{FigFitH218O}%
\end{figure}

The broad absorption observed on the blue part of the \waterhuit~and \watersept~para and ortho ground-state line profiles can give us some interesting information concerning the outer source structure. When we overplot these lines on the line of the main isotopolog (see Fig.\ref{FigCompare} and \ref{FigCompare2}), it first appears that \watersept~and \waterhuit~line profiles are identical (within the noise): the broad absorption seems to be present only in the blue part; this is what we assumed in Sect.\ref{sec:HIFIanalysis} for the Gaussian fitting. Conversely, the \water~profile differs with (weak) emission throughout most of the blue velocity interval ($\sim$ -72 to -53 \kms). From these observations we can first infer that this \watersept~and \waterhuit~material is dense and cold enough to absorb the warmer outflowing gas and that it is situated between the outflow and us. In adddition, the fact that the outflow is preferentially absorbed in its blue part most
likely reveals a cold outer gas envelope in expansion. Figure \ref{FigFitH218O} shows the Gaussian fitting for the \hoI~line when we assume that the absorption is rather centered on the source velocity: we obtain a component at -45 \kms~whose FWHM is 34 \kms.

\begin{figure}
\centering
\resizebox{\hsize}{!}{\includegraphics{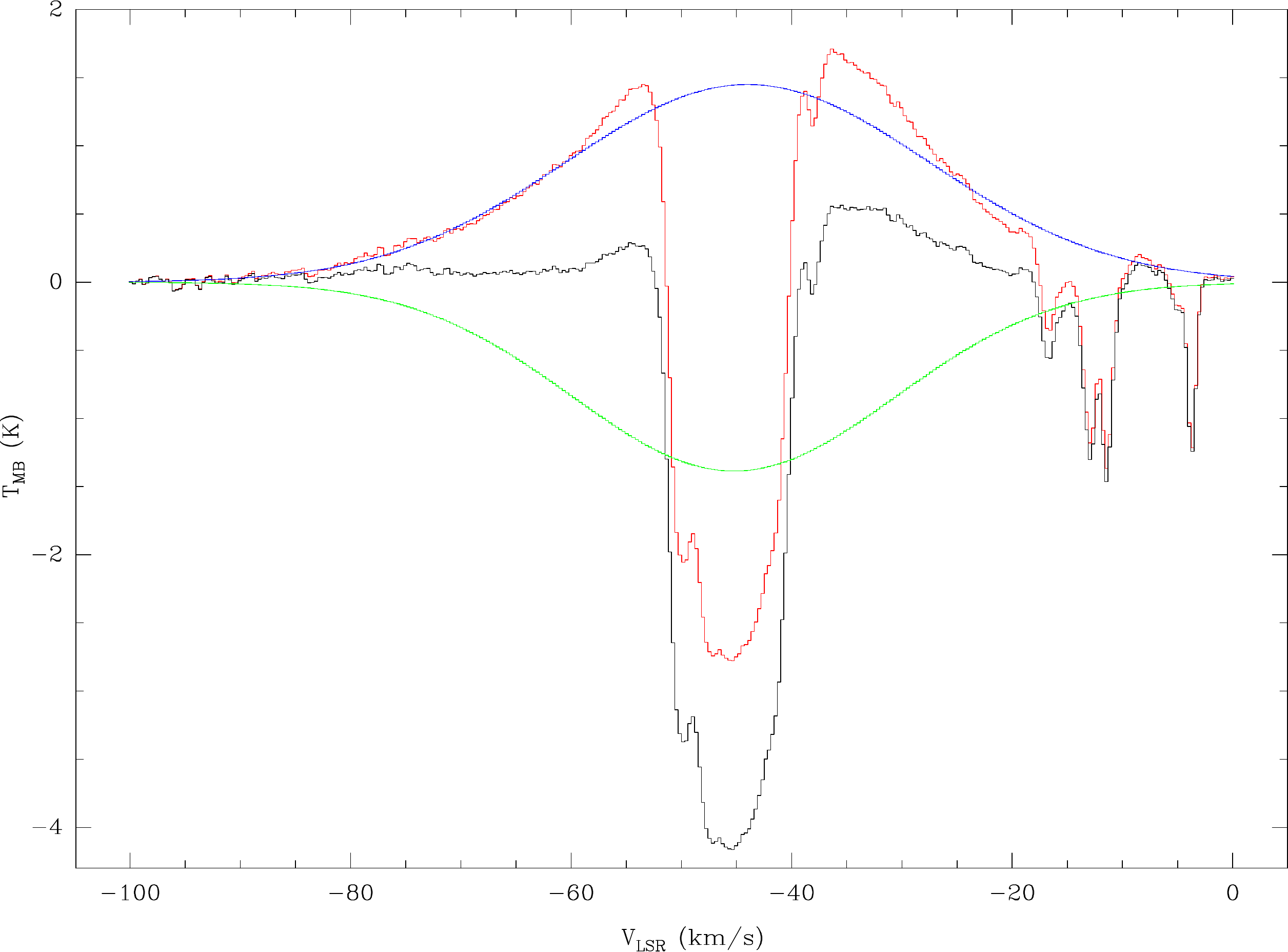}}
\caption{Resulting line profile (in red) of the \hoK~line corrected from the absorption shown in green and derived from the \hoI~Gaussian line fitting. The original spectra are shown in black. The blue curve is the Gaussian fitting of the outflow.}
\label{Fig1113Total}%
\end{figure}

Assuming the peak intensity ratio ($\sim$6.2) of the \water$/$\waterhuit~main narrow absorption features at -43.2 \kms (see Table \ref{table_param}) is valid for this broad absorption as well, we extrapolated the corresponding absorption in the \hoK~line profile. We then tried to remove this Gaussian absorption. The resulting water line profile is shown on Fig.\ref{Fig1113Total} and reveals a strong outflow centered at -44 \kms~whose FWHM is 39 \kms, broader than the first estimate we made in Sect.\ref{sec:HIFIanalysis}.

\begin{figure}
\centering
\resizebox{\hsize}{!}{\includegraphics{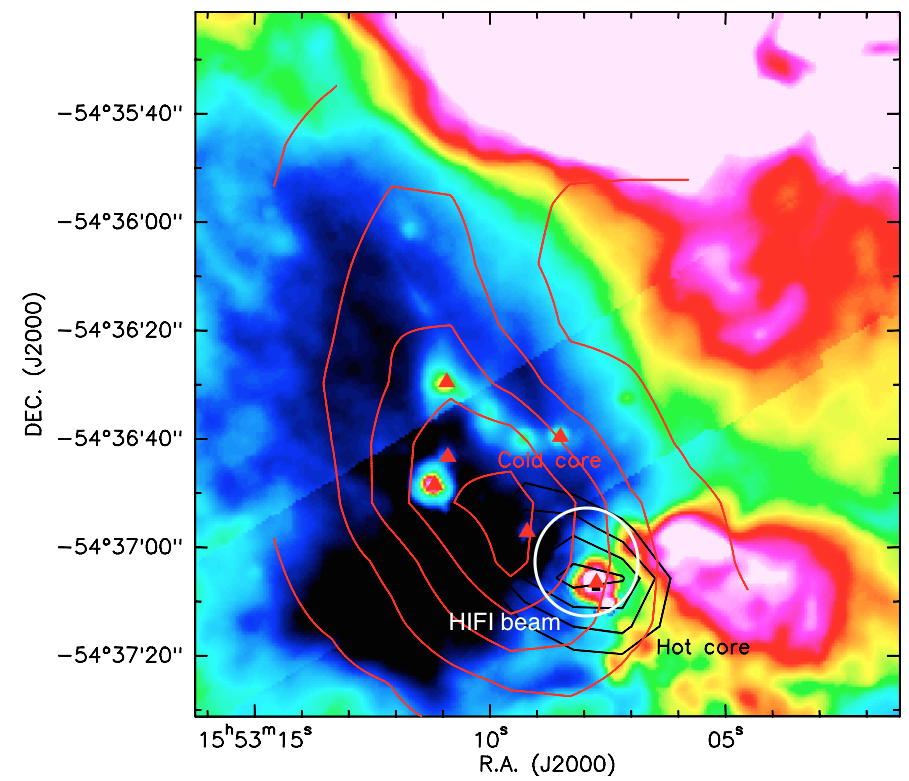}}
\caption{APEX images from \citet{2006A&A...454L..91W} in the N$_2$H$^+$ (3--2) and the CH$_3$OH lines (red and black contours) overlaid on the 8 $\mu$m GLIMPSE emission. Embedded GLIMPSE point sources with massive YSO characteristics are marked with triangles. The \textit{Herschel} 1113 GHz beam is indicated by the white circle centered on the HIFI observed position.}
\label{G327_wyr}%
\end{figure}

This cold absorbing material most likely corresponds to the cold clump found by \citet{2006A&A...454L..91W} in N$_2$H$^+$ and located 30$\arcsec$ (0.4 pc) northeast of the hot core, but extending well over our HIFI pointed observed area (see Fig.\ref{G327_wyr}). The velocity  of this cloud is -45.9 \kms~, comparable to what we deduce here. Of course the line width in N$_2$H$^+$ is narrower than in water as a result of excitation mechanisms. This cold absorbing material could also be compatible with an expanding shell on the HII region (see Fig. \ref{rv}) centered at -45 \kms. When we consider the distance between the HII region and SMM1 (110 arcsec=1.65 pc), the time needed to cross this distance from the center of the expanding HII to SMM1 (at a velocity of 6.5 \kms, see Sect. \ref{HIIregion}) would be 2.6 $10^5$ yr. 

\begin{figure}
\centering
\includegraphics[width=9.cm]{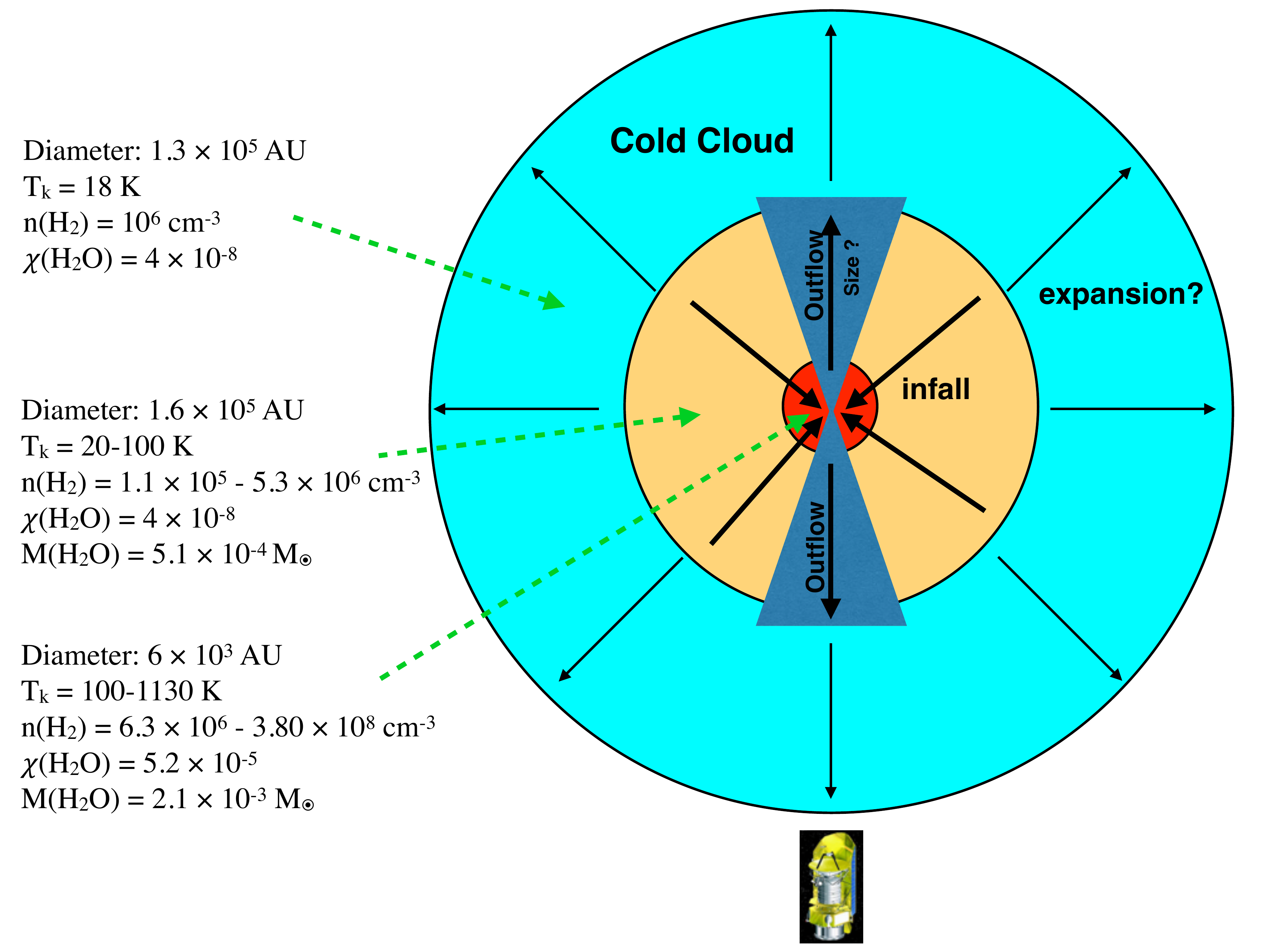}
\caption{Sketch of the G327 hot core source (arbitrary scale). See details in Sect. \ref{sec_disc}.}
\label{dessinG327}%
\end{figure}

 We used RADEX online\footnote{http://home.strw.leidenuniv.nl/$\sim$moldata/radex.html} to investigate the excitation conditions for the derived \water~broad absorption. In perfect agreement with \citet{2006A&A...454L..91W}, we deduce T$_{kin}=18$ K, n$_{\rm{H}_2}=10^6$ cm$^{-3}$, and N$_{\rm{H}_2}=10^{24}$ cm$^{-2}$ for the cold cloud ($\chi_{\rm{H_2O}}=4\times10^{-8}$ is assumed). If the density is constant over the cloud, the absorbing material should extend over $6.7\times10^4$ AU. Because the absorption is stronger for the blue part of the spectra, we propose that the outflow is seen face-on behind a cold envelope in expansion, as shown in Fig.\ref{dessinG327}.
  
 As explained in Sect.\ref{sec:HIFIanalysis}, the HIFI observed position is centered between SMM2 and SMM1, and so are maybe some of the physical components derived from our model. Interestingly, Fig. \ref{map987out} shows that the outflow is indeed rather centered away from the hot core, closer to the peak of the thermal CH$_3$OH emission detected by \citet{2006A&A...454L..91W} (see Fig. \ref{G327_wyr}) and might be associated to the class I methanol masers detected by \citet{voronkov2014} between SMM1 and SMM2. 

\subsection{Accretion rate and water content}

From the infall velocity estimated from our model (-3.2 \kms~as revealed by the 752 GHz \water~line), using n$_{H_2}=10^7$ cm$^{-3}$, we deduce a mass infall rate of $1-1.3\times 10^{-2}$ \Msol$/$yr. When we consider a mass of 20 M$_{\odot}$ within a radius of 100 R$_{\odot}$, this implies an accretion luminosity $L_{acc}\sim10^4$ \lsol~, which is high enough to overcome the radiation pressure that is due to the stellar luminosity (i.e., $\sim 3. \times10^{-4}$ \Msol$/$yr). Nevertheless, it is important to stress that this accretion luminosity is very sensitive to the assumed density and radius, and as a consequence, this comparison has to be cautiously considered.  The accretion rate is roughly three times greater than the rate derived by \citet{wyrowski2016} from the NH$_3$ $3_{2+}-2_{2-}$ line, but twice lower than the free-fall accretion (if we assume that the entire envelope mass is collapsing).
 We tried to estimate the size of the infall region as revealed by \hoE~line emission, but the line quickly vanishes out of the central source in the corresponding map. We derived a minimum size of the infall region of 20 $\arcsec$ to be compared with the size of the cool ($T_k=20-100$K) envelope, that is, 80$\arcsec$. Moreover, we did not find any evidence of rotation.

The inner water abundance ($5.2 \times 10^{-5}$) derived in Sect. \ref{sec_abun} for the hot core is slightly higher than what has been found for mid-IR-quiet massive protostars by \citet{herpin2016}. When we consider the high infall velocity we estimated for this source (-3.2 \kms), this value agrees with the scenario proposed by Herpin and collaborators that   higher inner abundance are observed for higher infall $  \text{or }$expansion velocities in the protostellar envelope.
We also estimate the amount of water in the inner (T$>$100 K) and outer regions to be $2.1\times10^{-3}$ and $5.1\times10^{-4}$ \Msol, respectively. This inner region that holds 80\% of the water corresponds to the compact area of 2$\arcsec$ where \citet{2000ApJ...545..309G} and \citet{bisschop2013} situated most of the organic species they observed, coming from the grain mantle evaporation.

\section{Conclusions}
\label{sec:Conclusions}

We have presented new \textit{Herschel}/PACS continuum maps at 89 and 179 $\mu$m that encompass the whole region (HII region and IRDC) and APEX$/$SABOCA map at 350 $\mu$m of the IRDC. These maps were combined with new spectral \textit{Herschel}$/$HIFI maps toward the IRDC region at 987 and 1113 GHz. In addition, we analyzed and modeled HIFI pointed observations of 15 water lines toward the G327 hot core region. 
 
Our data show that the distribution of the continuum emission at 89 and 179 $\mu$m follows the thermal continuum emission observed at larger wavelengths, with a peak at the position of the hot core and a secondary peak in the H{\sc ii} region, and an arch-like layer of hot gas west of this H{\sc ii} region. The same morphology is observed in the \hoK~line, in absorption toward all submillimeter dust condensations, while the 
$2_{02}-1_{11}$ line is seen in emission except at the positions of the hot core and of SMM2. We estimated column densities of $10^{15}$ and $2\times10^{14}$ cm$^{-2}$  at the hot core and IRDC position, respectively, corresponding to water abundances of 3$\times 10^{-8}$ in the outer envelope toward the hot core, while the abundance of water does not  change along the IRDC with values close to 10$^{-8}$. The water abundance is observed to be slightly larger in the more evolved object, that is, in the hot core, than in the IRDC, where no variation is seen. These values are also higher than what \citet{vandertak2010} derived in the DR21 region. The inner water abundance is estimated to
be $5.2 \times 10^{-5}$ for the hot core, in agreement with the
higher inner abundance for higher infall $  \text{or }$expansion velocities in the protostellar envelope \citep[][]{herpin2016}.

The map analysis combined with the radiative transfer modeling of the pointed spectral lines reveals a complex source structure of the hot core region. An outflow is detected, most likely seen face-on instead of centered away from the hot core, closer to the peak of the thermal CH$_3$OH emission, and it might be associated with the class I methanol masers between SMM1 and SMM2.  A strong infall associated with supersonic turbulence is also detected toward the hot core position at -3.2 \kms~(at $\sim$1500 AU), leading to an estimated mass infall rate of $1-1.3\times 10^{-2}$ \Msol$/$yr, which is high enough to overcome the radiation pressure that is due to the stellar luminosity. We derived a minimum size of the infall region of 20 $\arcsec$. No velocity gradient in the envelope can be inferred from the data, in contrast to what has been observed for the mini-starburst region W43-MM1 by \citet{jacq2016}. 

Moreover, we infer that a cold outer gas envelope in expansion is situated between the outflow and the observer, located 30$\arcsec$ (0.4 pc) northeast of the hot core, but extending  over $6.7\times10^4$ AU, hence somewhat comparable to W43-MM1. This cold absorbing material most likely corresponds to the cold clump found by \citet{2006A&A...454L..91W} in N$_2$H$^+$ , but it extends well beyond our HIFI pointed observed area. 

\begin{acknowledgements}

{\it Herschel} is an ESA space observatory with science instruments provided
by European-led Principal Investigator consortia and with important
participation from NASA. 
HIFI has been designed and built by a consortium of
 institutes and university departments from across Europe, Canada and the United States under the leadership of SRON Netherlands Institute for Space
 Research, Groningen, The Netherlands and with major contributions from
 Germany, France and the US. Consortium members are: Canada: CSA,
 U.Waterloo; France: CESR, LAB, LERMA, IRAM; Germany: KOSMA,
 MPIfR, MPS; Ireland, NUI Maynooth; Italy: ASI, IFSI-INAF, Osservatorio
 Astrofisico di Arcetri- INAF; Netherlands: SRON, TUD; Poland: CAMK, CBK; Spain: Observatorio Astron{\'o}mico Nacional (IGN), Centro de
 Astrobiolog{\'i}a
 (CSIC-INTA). Sweden: Chalmers University of Technology - MC2, RSS $\&$
 GARD; Onsala Space Observatory; Swedish National Space Board, Stockholm
 University - Stockholm Observatory; Switzerland: ETH Zurich, FHNW; USA:
 Caltech, JPL, NHSC.). 
\end{acknowledgements}

\begin{appendix}

\section{PACS line maps of the HII region}
\label{sec:PACS_HII}

\begin{figure*}
\centering
\includegraphics[width=8.6cm]{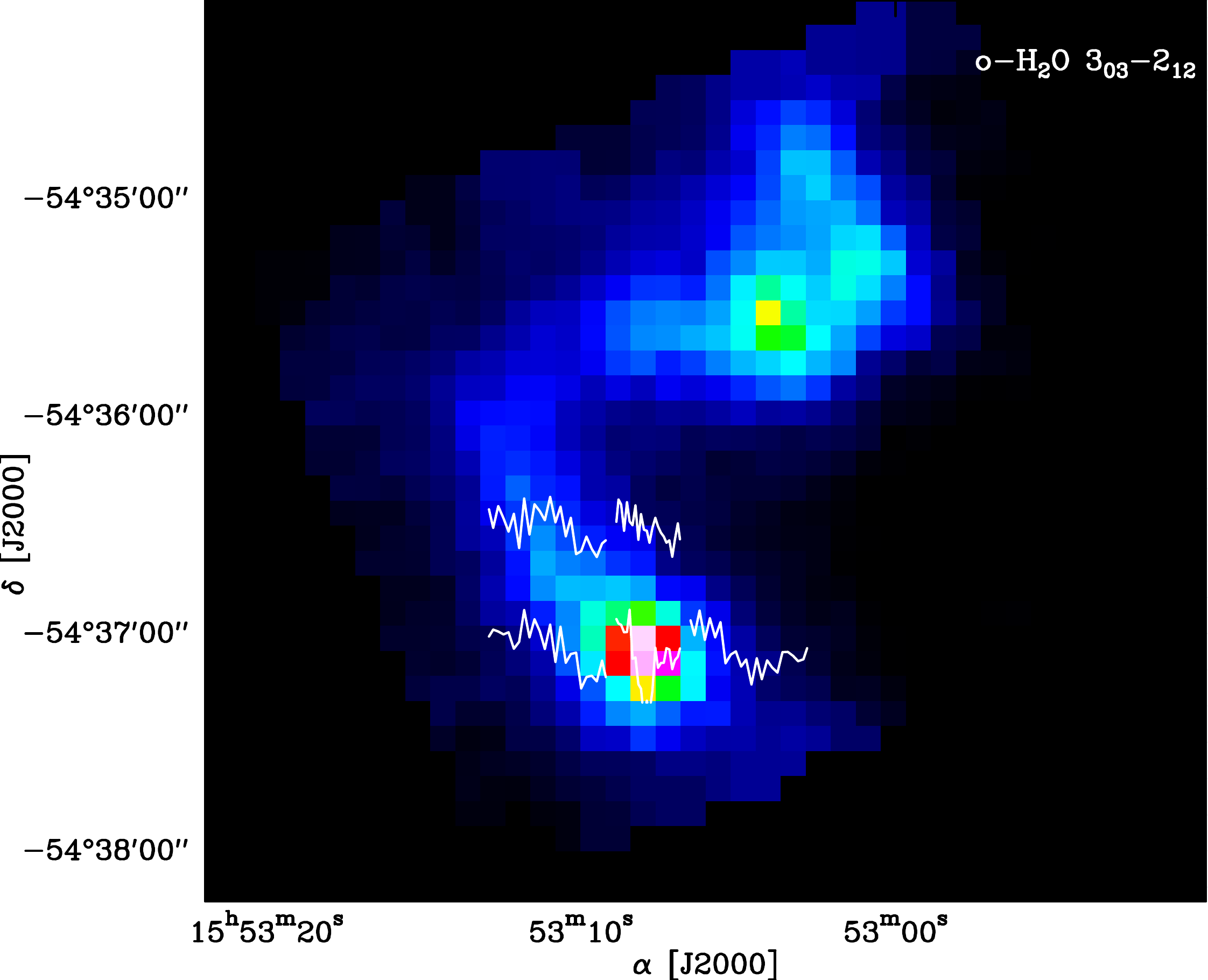}
\includegraphics[width=8.6cm]{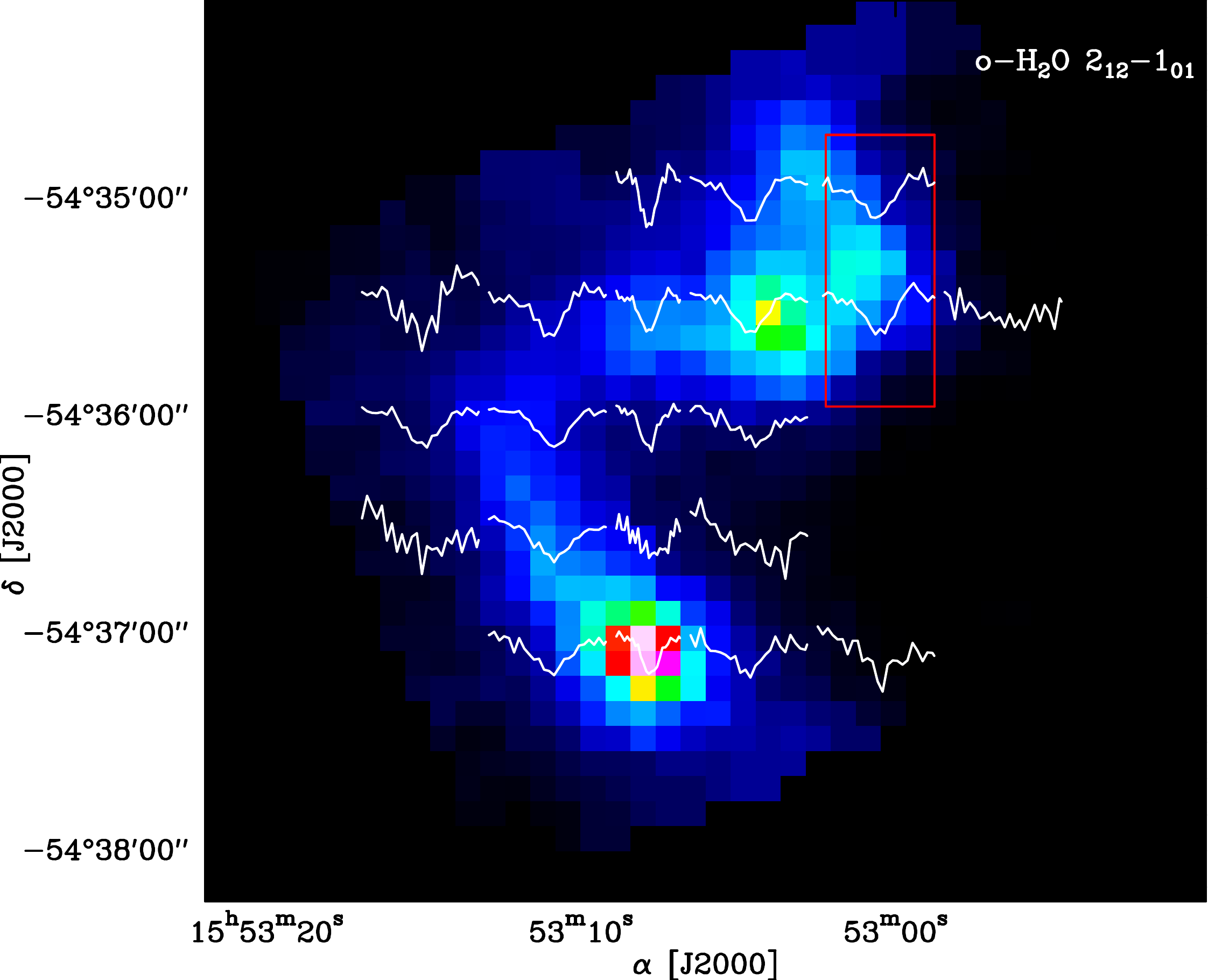}
\caption{Spectral map of the $o$-H$_2$O $2_{12}-1_{01}$  ({\em left panel}) and $o$-H$_2$O $3_{03}-2_{12}$ ({\em right panel}) lines overlaid on the PACS continuum emission at 179\,$\mu$m. The spectra are displayed in the velocity range [-350,250] km\,s$^{-1}$ as line-to-continuum ratio. In the left panel, the red rectangle outlines the region where the CH$^+$ (2-1) line is detected (emission at redshifted velocities compared to the $o$-H$_2$O $2_{12}-1_{01}$ transition).}
\label{fig:PACS_HII_179}
\end{figure*}

\end{appendix}

\end{document}